\documentclass[11pt,a4paper]{article}
\usepackage{jheppub}
\usepackage{latexsym}
\usepackage{pdflscape}
\usepackage{bm}

\def\cob{\delta}

\DeclareMathOperator{\Tr}{Tr}

\newcommand{\hf}{\frac{1}{2}}
\newcommand{\qu}{\frac{1}{4}}

\def\si{\sigma}

\def\del{\partial}

\def\bra{\langle}
\def\ket{\rangle}
\def\lf{\left}
\def\ri{\right}

\def\h#1{\widehat{#1}}

\def\ga{\gamma}
\def\Ga{\Gamma}

\def\rt#1{\sqrt{#1}}
\def\sitarel#1#2{\mathrel{\mathop{\kern0pt #1}\limits_{#2}}}

\newcommand{\nn}{\nonumber \\}

\title{Probing non-perturbative effects in M-theory on orientifolds}

\author{Kazumi Okuyama}

\affiliation{Department of Physics, Shinshu University, Matsumoto 390-8621, Japan}

\emailAdd{kazumi@azusa.shinshu-u.ac.jp}

\abstract{
Using holography, we study non-perturbative effects in M-theory
on orientifolds
from the analysis of the $S^3$ partition functions
of dual field theories.
We consider the $S^3$ 
partition functions of
$\mathcal{N}=4$ 
Yang-Mills theory with $O(n)$ gauge symmetry 
coupled to one (anti)symmetric and $N_f$
fundamental hypermultiplets
from the Fermi gas approach.
In addition to the
worldsheet instanton and membrane instanton
corrections to the grand potential,
which are also present in the $U(n)$ Yang-Mills case,
we find that there exist ``half instanton''
corrections
coming from the effect of orientifold plane. 
}

\begin{document}
\maketitle

\section{Introduction}
\label{sec:intro}
In the last few years, we have witnessed
a tremendous progress in our understanding
of non-perturbative effects in M-theory.
In particular,
in the case of $\mathcal{N}=6$
$U(N)_k\times U(N)_{-k}$
ABJM theory,
which is holographically dual to
M-theory on $AdS_4\times S^7/\mathbb{Z}_k$,
we have a complete understanding
of non-perturbative corrections
thanks to the relation to topological string
on local $\mathbb{P}^1\times\mathbb{P}^1$ \cite{HMMO}
(see \cite{Hatsuda:2015gca} for a review).
Especially, the grand partition functions of ABJM theory
at $k=1,2,4$ are completely determined in closed forms \cite{Codesido:2014oua,Grassi:2014uua}.

However, for lower supersymmetric theories
we still do not have a
detailed understanding of non-perturbative effects
in M-theory.
$d=3~\mathcal{N}=4$ theories are particularly
interesting since these theories
are sometimes related by mirror symmetry
exchanging the Higgs branches and Coulomb branches \cite{Intriligator:1996ex}. 
Such $\mathcal{N}=4$ theories
naturally appear as the worldvolume theories
on M2-branes probing ALE singularities.
For instance, $N$ M2-branes probing $A_{N_f-1}$
ALE singularities have two descriptions related by
mirror symmetry:  $U(N)$ Yang-Mills theory
 with one adjoint and $N_f$ fundamental
 hypermultiplets, and a $\h{A}_{N_f-1}$ quiver gauge
 theory \cite{Intriligator:1996ex, deBoer:1996mp}.
The former theory appears as a worldvolume theory on $N$ 
D2-branes in the presence of $N_f$ D6-branes, and the latter
description comes from the 
M-theory lift of D6-branes as Taub-NUT space.
 In the large $N$ limit, these theories are
 holographically dual to M-theory
 on $AdS_4\times S^7/\mathbb{Z}_{N_f}$,
 and we can study non-perturbative effects in this background
 from the $S^3$ partition function of the field theory side.
 In \cite{Mezei:2013gqa,GM,Hatsuda:2014vsa},
 the $S^3$ partition function of these theories,
 known as the $N_f$ matrix model,
 are studied using the Fermi gas formalism \cite{MP2}.
 It turned out that the instanton corrections
 in the $N_f$ matrix model are quite different from those
 in the ABJM theory and have a very intricate structure \cite{Hatsuda:2014vsa}.
Instanton corrections in more general $\mathcal{N}=4$
quiver gauge theories are also studied in \cite{Moriyama:2014waa,Moriyama:2014nca,Hatsuda:2015lpa},
but it is fair to say that we are far from the complete picture.

In this paper, we will study a natural generalization of $N_f$ matrix model:
$S^3$ partition
functions of $d=3~\mathcal{N}=4$ $O(n)$ or $USp(n)$
Yang-Mills theories
with $N_f$ fundamental and one (anti)symmetric hypermultiplets, 
first studied in \cite{Mezei:2013gqa} using the Fermi gas approach.
In the Type IIA brane constructions, 
such models appear as worldvolume theories on
$N$ D2-branes in the presence of $N_f$ D6-branes and a orientifold plane.
We will denote the model of gauge group $G$
with $N_f$ fundamental and one symmetric (or anti-symmetric) hypermultiplets
as $G+S$ (or $G+A$), respectively.

By mirror symmetry,
the $USp(2N)+A$ model is
dual to a $\h{D}_{N_f}$
quiver gauge theory\footnote{The Fermi gas formalism of
$\h{D}$-type quiver gauge theories has appeared in
\cite{Assel:2015hsa,Moriyama:2015jsa}.}
\cite{Intriligator:1996ex,deBoer:1996mp},
which can be interpreted as the worldvolume theory
on M2-branes probing
the $D_{N_f}$ ALE singularity.
In the large $N$ limit, this theory is holographically dual 
to M-theory on $AdS_4\times S^7/\Ga_{D_{N_f}}$
where $\Ga_{D_{N_f}}$ is the dihedral subgroup of $SU(2)$.
This opens up an avenue to study M-theory on orientifolds from
the analysis of $S^3$ partition
functions of dual field theories. 
In particular, we can study
the effects of orientifold plane
in the M-theoretic regime where the string coupling $g_s$
of Type IIA theory becomes large.
Orientifolds in Type IIB theory 
can be described in F-theory,
while the strong coupling behavior of Type IIA orientifolds
is still poorly understood.
Our work is a first step towards
the understanding of non-perturbative effects
in M-theory on orientifolds\footnote{In a recent paper \cite{Moriyama:2015asx},
the orientifold ABJM theory is studied from the Fermi gas approach.}.

We find that the $USp(n)+A$(or $S$) model
is related to the $O(n)+A$(or $S$) model
by a shift of $N_f$, hence it is sufficient to consider the 
$O(n)$ case only.
For the $O(n)+A$ model
we find that there are three types of instantons:
worldsheet instantons, membrane instantons, and ``half instantons''.
The first two types have 
direct analogues in the $N_f$ matrix model, while
the last type is a new one coming from the effect of orientifold plane.
In the Fermi gas picture,
orientifolding corresponds to
the reflection of
fermion coordinate $x\to -x$,
and the ``half instantons'' can be naturally identified
as the contribution of the twisted sector of this reflection. 
We find that the sign of this contribution 
depends on the parity $(-1)^n$ of the gauge group
$O(n)$.
On the other hand, we could not find a clear picture
of the instanton corrections in the $O(n)+S$ model.

This paper is organized as follows:

In section \ref{sec:fermigas},
we first review the
Fermi gas formalism of the $S^3$
partition functions of $G+A$ or $G+S$ models \cite{Mezei:2013gqa}.
Then we explain our algorithm 
to compute the partition functions
of these models exactly.

In section \ref{se:pert}, we determine
the coefficients $C,B$ and $A$
in the perturbative part of grand potential \eqref{Jpert}.
The results are summarized in Table \ref{table:ABC}.

In section \ref{sec:non-pert},
we study the non-perturbative corrections
to the grand potential using our
data of exact partition functions.
For the $O(n)+A$ model,
we find the 
first few coefficients of instanton corrections as a function of
$N_f$.
We also comment on the instanton corrections in the $O(n)+S$ model.

In section \ref{sec:WKB1},
we compute the WKB expansion
of grand potential using the density matrix operator in
\cite{Assel:2015hsa},
and reproduce the
coefficients $C,B$ and $A$
for the $O(2N+1)+A$ model.

In section \ref{sec:WKB2},
using the different form of operator in \cite{Mezei:2013gqa},
we compute the WKB expansion
of the ``twisted spectral trace''
defined in \eqref{BarnesJR}.
We argue that this contribution
is related to the effect of orientifold plane.

Finally, we conclude in section \ref{sec:conclusion}.
Additionally, we have two Appendices \ref{se:Jnp} and \ref{se:Wigner}.
In Appendix \ref{se:Jnp}, we summarize
the non-perturbative part of grand potential
$J_\text{np}(\mu)$ for various (half-)integer $N_f$,
determined from our data of exact partition functions.
In Appendix \ref{se:Wigner},
we explain the derivation of the Wigner transform in \eqref{rhoDW}.

\section{Fermi gas formalism and exact computation of partition functions}
\label{sec:fermigas}

We study the $S^3$
partition functions of 
$\mathcal{N}=4$ $G+A$ and $G+S$ models with
$G=O(n)$ or $USp(n)$,
considered previously in \cite{Mezei:2013gqa}.
Such models naturally appear as worldvolume theories on D2-branes in the presence of 
$N_f$ D6-branes and a orientifold plane.

As discussed in \cite{Mezei:2013gqa},
depending on the type of orientifold plane,
we find the following
models as worldvolume theories on  D2-branes:
\begin{itemize}
 \item $O(2N)+A$: $N$ D2-branes and $N_f$ D6-branes with a O2$^{-}$ plane.
\item $O(2N+1)+A$: $N$ D2-branes and $N_f$ D6-branes with a O2$^{-}$ plane on which
a half D2-brane got stuck.
\item $O(2N)+S$: $N$ D2-branes and $N_f$ D6-branes with a O6$^{+}$ plane.
\item $O(2N+1)+S$: $N$ D2-branes and $N_f$ D6-branes with a O6$^{+}$ plane on which
a half D2-brane got stuck.
\item $USp(2N)+A$: $N$ D2-branes and $N_f$ D6-branes with a O6$^{-}$ plane.
\item $USp(2N)+S$: $N$ D2-branes and $N_f$ D6-branes with a O2$^{+}$ plane.
\end{itemize}
To preserve $\mathcal{N}=4$ supersymmetry,
we consider a configuration of D2-branes and O2-planes extending  
in the directions $(x^0,x^1,x^2)$,
and D6-branes and O6-planes
extending  
in the directions $(x^0,x^1,\cdots,x^5)$
\cite{Mezei:2013gqa}.
They share the common three dimensional spacetime
$(x^0,x^1,x^2)$
on which the above $d=3~\mathcal{N}=4$ theories live. 

In \cite{Mezei:2013gqa}, it is found that the $S^3$ partition functions of above 
models can be written as
a system of $N$ fermions in one-dimension ($x_i\in\mathbb{R}$)
\begin{equation}
\begin{aligned}
Z(N,N_f)&=\frac{1}{N!}\int\frac{d^Nx}{(4\pi)^N} \prod_{i=1}^N \frac{(2\sinh \frac{x_i}{2})^{2a}
(2\sinh x_i)^{2b}(2\cosh x_i)^{1-2d}}{(2\cosh\frac{x_i}{2})^{2N_f+2c}}
\det\left(\frac{1}{2\cosh x_i+2\cosh x_j}\right) \\
&=\frac{1}{N!}\sum_{\si\in S_N}(-1)^\si \int d^Nx
\prod_{i=1}^N\rho(x_i,x_{\si(i)}),
\end{aligned} 
\label{Zint}
\end{equation}
where the density matrix $\rho(x,y)$ is given by
\begin{equation}
\begin{aligned}
  \rho(x,y)&=\frac{\rt{V(x)V(y)}}{2\cosh x+2\cosh y},\\
V(x)&=\frac{1}{4\pi} \frac{(2\sinh \frac{x}{2})^{2a}
(2\sinh x)^{2b}(2\cosh x)^{1-2d}}{(2\cosh\frac{x}{2})^{2N_f+2c}}.
\end{aligned}
\label{rhoxy}
\end{equation}
The parameters $a,b,c,d$ for each model
are summarized in Table \ref{abcd}.
\begin{table}[h]
\begin{center}
\begin{tabular}[t]{|l|c|c|c|c|}
\hline
model & $a$ & $b$ & $c$& $d$\\\hline
$O(2N)+A$ & 0& 0& 0& 0\\ \hline
$O(2N+1)+A$ & 1& 0& 1& 0 \\ \hline
$O(2N)+S$ & 0& 0& 0& 1\\ \hline
$O(2N+1)+S$ & 1& 0& 1& 1 \\ \hline
$USp(2N)+A$ & 0& 1& 0& 0  \\ \hline
$USp(2N)+S$ & 0& 1& 0& 1 \\ \hline
\end{tabular} 
\end{center}
\caption{$a,b,c,d$ for various models}
\label{abcd}
\end{table}
In \eqref{Zint}, we have fixed
the overall normalization of $Z(N,N_f)$ in such a way that $Z(N=0,N_f)=1$,
which is a natural normalization in the Fermi gas formalism \cite{MP2}.
Note that our normalization of $Z(N,N_f)$
is different from \cite{Mezei:2013gqa}\footnote{One might think that there is still
an ambiguity to change the normalization
$Z(N,N_f)\to c^N Z(N,N_f)$,
 with some positive constant $c$,
 which is equivalent to a shift of chemical potential $\mu\to\mu+\log c$.
However, there is no room for this change of normalization
since a shift of chemical potential will spoil the absence of $\mu^2$
term in the perturbative part of grand potential \eqref{Jpert}.}.
As discussed in \cite{MP2},
to study the 
non-perturbative corrections,
it is more convenient to
consider the grand partition function
by summing over $N$ with fugacity $e^\mu$
\begin{equation}
  \Xi(\mu)=\sum_{N=0}^\infty Z(N,N_f)e^{N\mu}.
  \label{Xidef}
\end{equation} 
From \eqref{Zint}, one can show that 
$\Xi(\mu)$
can be written as
a Fredholm determinant of the density matrix $\rho$
\begin{equation}
 \Xi(\mu)=\text{Det}(1+e^\mu\rho).
\end{equation}
More physically, $\rho$ is identified with the Hamiltonian
$H$
of the fermion system as
\begin{equation}
 \rho=e^{-H}.
\end{equation}
In the following sections, we will study the large $\mu$
expansion of the grand potential $J(\mu)$
\begin{equation}
 J(\mu)=\log\Xi(\mu)=\sum_{\ell=1}^\infty \frac{(-1)^{\ell-1}e^{\ell\mu}}{\ell}
\Tr(\rho^\ell).
\label{granddef}
\end{equation}

From \eqref{Zint} and Table \ref{abcd}, one can easily see that
the partition functions of $USp(2N)$ theory
and $O(2N+1)$ theory are related by
a shift of $N_f$
\begin{equation}
\begin{aligned}
 Z(N,N_f)_{USp(2N)+A}&= Z(N,N_f-2)_{O(2N+1)+A},\\
Z(N,N_f)_{USp(2N)+S}&=Z(N,N_f-2)_{O(2N+1)+S}.
\label{O-Sp-relation}
\end{aligned} 
\end{equation}
Therefore, for our purposes
it is sufficient to consider
the models with $O(n)$ gauge group.

We can compute the canonical partition function $Z(N,N_f)$ at fixed $N$
once we know the trace $\Tr\rho^\ell$ from $\ell=1$ to $\ell=N$. 
Using the Tracy-Widom lemma \cite{TW},
the $\ell^\text{th}$ power of $\rho$ can be systematically computed
by constructing a sequence of functions $\phi_\ell(x)~(\ell=0,1,2,\cdots)$
\begin{equation}
 \begin{aligned}
 \rho^\ell(x,y)&=\frac{\rt{V(x)V(y)}}{2\cosh x+(-1)^{\ell-1}2\cosh y}
\sum_{j=0}^{\ell-1}(-1)^j \phi_j(x)\phi_{\ell-1-j}(y),\\
\phi_\ell(x)&=\frac{1}{\rt{V(x)}}\int_{-\infty}^\infty dy\, \rho(x,y)\rt{V(y)}\phi_{\ell-1}(y),\quad\phi_0(x)=1.
\end{aligned}
\label{TW-rho}
\end{equation}
Then $\Tr\rho^\ell$ is given by
\begin{equation}
 \begin{aligned}
  \Tr\rho^{2n}&=\int_{-\infty}^\infty  dx 
\frac{V(x)}{2\sinh x}
\sum_{j=0}^{2n-1}(-1)^j \frac{d\phi_j(x)}{dx}\phi_{2n-1-j}(x),\\
\Tr\rho^{2n+1}&=\int_{-\infty}^\infty  dx 
\frac{V(x)}{4\cosh x}
\sum_{j=0}^{2n}(-1)^j \phi_j(x)\phi_{2n-j}(x).
 \end{aligned}
\label{TW-tr}
\end{equation}
The integrals  in \eqref{TW-rho} and \eqref{TW-tr}
can be easily evaluated by
rewriting them as contour integrals and picking up residues, as in the case of ABJM
theory \cite{PY,HMO2}.
Using this algorithm, we have computed
the exact values of partition functions 
$Z(N,N_f)$ of our models
for various integer $N_f$ and half-integer $N_f$ 
up to some high $N=N_\text{max}$, where $N_\text{max}$
is about 20-30.\footnote{
The data of exact values of $Z(N,N_f)$
are attached as ancillary files
to the {\tt arXiv} submission of this paper.}
Note that for a physical theory $N_f$ should be an integer,
but at the level of matrix model
\eqref{Zint} we can consider analytic continuation
of $N_f$ to arbitrary continuous values. 
Such analytic continuation in $N_f$ is
implicitly assumed in what follows.

Before moving on, let us comment on some interesting
relations between our models \eqref{Zint}
and some other theories.
First, by mirror symmetry of $d=3$ $\mathcal{N}=4$
theories, the $USp(2N)+A$ model
is dual to a $\h{D}_{N_f}$
quiver gauge theory with one fundamental flavor node added \cite{deBoer:1996mp}.
The equivalence of the $S^3$ partition functions of these two theories
can be shown by using the result of \cite{Assel:2015hsa}\footnote{We are grateful to Masazumi Honda for discussion on this point.}.

Second, we find a nontrivial relation
between the $USp(2N)+S$ model with $N_f=1$
and the ABJ theory with gauge group $U(N)_4\times U(N+1)_{-4}$
\begin{align}
Z_{USp(2N)+S}(N,N_f=1)&=Z_{\text{ABJ}}(N,k=4,M=1),
\label{Sp-ABJ}
\end{align}
where $M=N_2-N_1$ denotes the difference of the rank of gauge group $U(N_1)_k\times U(N_2)_{-k}$
of ABJ theory.
This relation \eqref{Sp-ABJ}
can be understood from the relation found in \cite{Grassi:2014uua}
\begin{align}
 \Xi_{\text{ABJ}}(\mu,k=4,M=1)=\Xi_{\text{ABJM}}^{-}(\mu,k=2),
\end{align}
where $\Xi_{\text{ABJM}}^{-}(\mu,k=2)$ is
the grand partition function of ABJM theory at $k=2$
computed from the odd-part $\rho_-$
of density matrix
\begin{equation}
 \rho_-(x,y)=\frac{\rho(x,y)-\rho(x,-y)}{2}.
\end{equation} 
One can easily show that the density matrix of $USp(2N)+S$ model with $N_f=1$
and the odd-part $\rho_-$  of ABJM theory at $k=2$
are equivalent, up to a rescaling  $x,y\to 2x,2y$ and 
a similarity transformation, hence the relation \eqref{Sp-ABJ}
follows.

Finally, we also find
the equivalence of the partition functions of 
$USp(2N)+A$ with $N_f=3$
and the $U(N)$ Yang-Mills theory
with one adjoint and $N_f$ fundamental hypermultiplets (the
$N_f$ matrix model)
with $N_f=4$
\begin{align}
 Z_{USp(2N)+A}(N,N_f=3)=Z_{U(N)+\text{adj}}(N,N_f=4).
\label{D3A3}
\end{align}
This is expected from the isomorphism
$D_3= A_3$. 
This relation \eqref{D3A3}
is recently proved in \cite{Honda}
using the technique in \cite{Assel:2015hsa}.

\section{Perturbative part}
\label{se:pert}
In this section, we consider the
large $\mu$ expansion of the grand potential \eqref{granddef},
which
takes the following form
\begin{equation}
\begin{aligned}
 J(\mu)&=J_\text{pert}(\mu)+J_\text{np}(\mu),\\
J_\text{pert}(\mu)&=\frac{C\mu^3}{3}+B\mu+A.
\end{aligned}
\label{Jpert}
\end{equation}
Here $J_\text{pert}(\mu)$ in \eqref{Jpert}
is called the perturbative pert of grand potential.
On the other hand, $J_\text{np}(\mu)$ in \eqref{Jpert}
represents the non-perturbative corrections
which are exponentially suppressed
in the large $\mu$ limit.
We will study $J_\text{np}(\mu)$ in the next section.

In the large $N$ limit,
the free energy $F=-\log Z(N,N_f)$
is approximated by the Legendre transform
of $J_\text{pert}(\mu)$
\begin{equation}
 F\approx N\mu_*-J_\text{pert}(\mu_*)\approx \frac{2}{3}C^{-\hf}N^{\frac{3}{2}},\quad(N\gg1),
\label{largeN-F}
\end{equation}
where $\mu_*$ is the saddle point value of the chemical potential
\begin{equation}
 \mu_*=\rt{\frac{N}{C}}.
\label{mu-star}
\end{equation}
Since the free energy on $S^3$ is a nice measure of the degrees of freedom
in $d=3$ theories \cite{Jafferis:2011zi},
\eqref{largeN-F} implies that the degrees of freedom
of our models scale as $N^{3/2}$, which is the expected behavior of M2-brane
theories \cite{KT}.

We would like to determine
the coefficients $C,B$ and $A$ in \eqref{Jpert} as a function of $N_f$.
The coefficient $C$ is already found in \cite{Mezei:2013gqa}
from the analysis of the classical Fermi surface.
The coefficient $B$ is a bit difficult since $B$
receives a correction
in the semi-classical WKB expansion (small-$\hbar$  expansion).
The coefficient $A$ is much harder to determine
since $A$ receives corrections from all orders in
the WKB expansion.

To circumvent this problem, we determine
the coefficients $B$ and $A$
by matching our exact values of $Z(N,N_f)$
and the perturbative partition function $Z_\text{pert}(N,N_f)$ given by the Airy function
\cite{FHM,MP2}
\begin{equation}
 \begin{aligned}
  Z(N,N_f)&=Z_\text{pert}(N,N_f)+Z_\text{np}(N,N_f),\\
Z_\text{pert}(N,N_f)&=\int_{\mathcal{C}}
\frac{d\mu}{2\pi i}e^{J_\text{pert}(\mu)-N\mu}=C^{-\frac{1}{3}}e^A\text{Ai}\Bigl[C^{-\frac{1}{3}}(N-B)\Bigr],
 \end{aligned}
 \label{Zpert}
\end{equation}
where $\mathcal{C}$ is a contour in the $\mu$-plane from $e^{-\frac{\pi i}{3}}\infty$
to $e^{\frac{\pi i}{3}}\infty$, and
$Z_\text{np}(N,N_f)$ denotes the non-perturbative corrections
coming from $J_\text{np}(\mu)$.
When $N$ becomes large, the non-perturbative corrections $Z_\text{np}(N,N_f)$
are highly suppressed, so
we can approximate the partition function by its perturbative part
$Z_\text{pert}(N,N_f)$.
By comparing the exact values of $Z(N,N_f)$ 
and $Z_\text{pert}(N,N_f)$ in \eqref{Zpert}, we find the coefficients 
$B$ and $A$
for various models, which are summarized in Table \ref{table:ABC}.
\begin{table}[th]
\begin{center}
\begin{tabular}[t]{|l|l|l|l|}
\hline
model & $C$ & $B$ & $A$ \\\hline
$O(2N)+A$ & $ \frac{1}{2\pi^2 N_f}$ & $\frac{1}{8 N_f}-\frac{N_f-1}{8}$ & $\qu A_c(2N_f)
+\frac{2N_f^2+7N_f+7}{2}A_c(1)-\frac{4N_f+5}{4}\log2$\\ \hline
$O(2N+1)+A$ &  $ \frac{1}{2\pi^2 N_f}$ &  $\frac{1}{8 N_f}-\frac{N_f+3}{8}$
& $\qu A_c(2N_f)
+\frac{2N_f^2+7N_f+7}{2}A_c(1)-\frac{1}{4}\log2$ \\ \hline
$O(2N)+S$ &  $ \frac{1}{2\pi^2 (N_f+2)}$& $\frac{1}{8 (N_f+2)}-\frac{N_f-1}{8}$
& $\qu A_c(2N_f+4)
+\frac{2N_f^2+N_f+1}{2}A_c(1)-\frac{3}{4}\log2$\\ \hline
$O(2N+1)+S$ &  $\frac{1}{2\pi^2 (N_f+2)}$& $\frac{1}{8 (N_f+2)}-\frac{N_f+3}{8}$
&$\qu A_c(2N_f+4)
+\frac{2N_f^2+N_f+1}{2}A_c(1)+\frac{4N_f+5}{4}\log2$\\ \hline
\end{tabular} 
\end{center}
\caption{The coefficients $C,B$ and $A$ in $J_\text{pert}(\mu)$.
$C$ is already found in \cite{Mezei:2013gqa}.}
\label{table:ABC}
\end{table}
As pointed out in \cite{Assel:2015hsa}, the computation of
$B$ in \cite{Mezei:2013gqa} has an error, and our results of $B$ are different from
\cite{Mezei:2013gqa}.
$A_c(k)$ in Table \ref{table:ABC} is the constant term in the grand potential of
$U(N)_k\times U(N)_{-k}$
ABJM theory, which is closely related to 
a certain resummation of the constant map contribution of topological string
\cite{Hatsuda:2014vsa,KEK,Hatsuda:2015owa}
\begin{equation}
  A_c(k)
=-\frac{k^2\zeta(3)}{8\pi^2}+4\int_0^\infty dx\frac{x}{e^{2\pi x}-1}
\log\left(2\sinh\frac{2\pi x}{k}\right).
\end{equation}
For integer $k$, $A_c(k)$ can be written in a closed form
\begin{equation}
 \begin{aligned}
  A_c(k)=\begin{cases}
\displaystyle
-\frac{\zeta(3)}{\pi^2 k}-\frac{2}{k} \sum_{m=1}^{\frac{k}{2}-1} m\Big(\frac{k}{2}-m\Big) \log \Big( 2 \sin \frac{2\pi m}{k} \Big) &(\text{even } k), \\
\displaystyle
-\frac{\zeta(3)}{8\pi^2 k}+\frac{k}{4}\log 2-\frac{1}{k} \sum_{m=1}^{k-1} g_m(k)(k-g_m(k)) \log \Big(2 \sin \frac{\pi m}{k} \Big) \; &(\text{odd } k).
\end{cases}
 \end{aligned}
\label{Acintk}
\end{equation}
where
\begin{equation}
g_m(k)=\frac{k+(-1)^m(2m-k)}{4}. 
\end{equation}
In particular, $A_c(1)$ appearing in Table \ref{table:ABC}
is given by
\begin{equation}
 A_c(1)=-\frac{\zeta(3)}{8\pi^2}+\qu\log2.
\end{equation}
\begin{figure}[tb]
\begin{center}
\begin{tabular}{cc}
\hspace{-4mm}
\includegraphics[width=7cm]{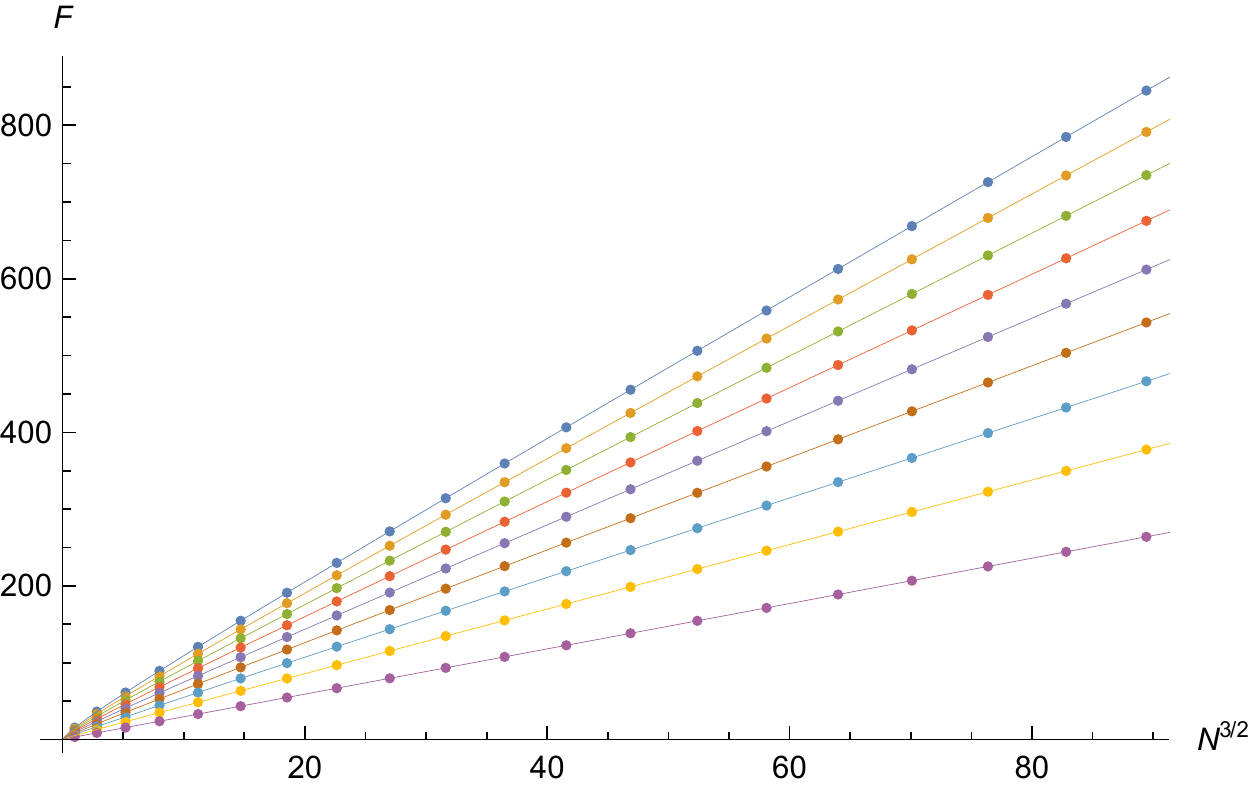}
\hspace{10mm}
&
\includegraphics[width=7cm]{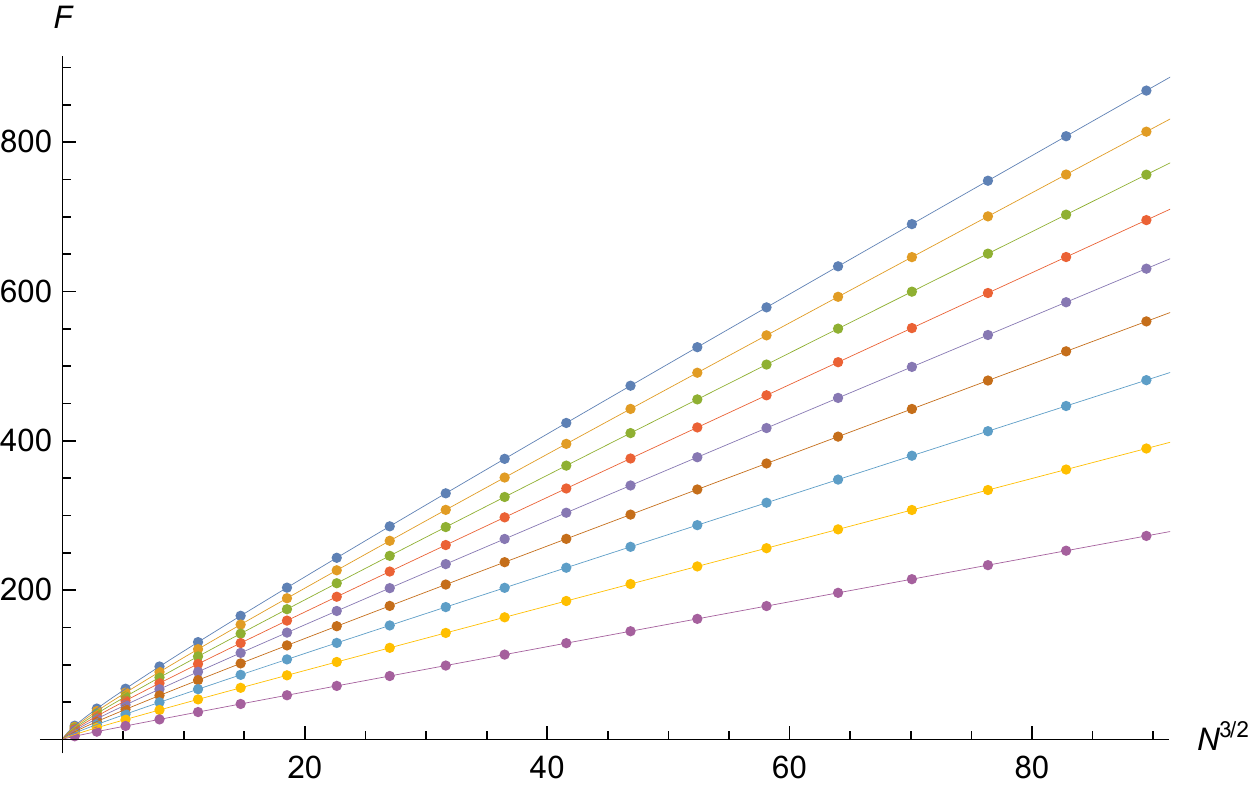}\\
(a) free energy of $O(2N)+A$ model
\hspace{10mm}
&
(b)  free energy of $O(2N+1)+A$ model
\end{tabular}
\end{center}
  \caption{We show  the plot of free energy $F=-\log Z(N,N_f)$ 
of (a) $O(2N)+A$ model and (b)
$O(2N+1)+A$ model for $N_f=1,2,\cdots,9$.
Note that the horizontal axis is $N^{3/2}$.
In both figures (a) and (b), $N_f$ increases from the bottom curve $(N_f=1)$
to the top curve $(N_f=9)$.
The dots in the figures are the 
exact values of free energy at integer $N$,
while
the solid curves represent 
the perturbative free energy
$F_\text{pert}=-\log Z_\text{pert}(N,N_f)$ given by the Airy function \eqref{Zpert}
with $C,B$ and $A$  in Table \ref{table:ABC}.
}
  \label{fig:Zpert}
\end{figure}
In Figure \ref{fig:Zpert}, 
we show the plot of free energy for the $O(n)+A$ models.
As we can see, the exact values of free energy
at integer $N$ exhibit a nice agreement with
the perturbative
free energy \eqref{Zpert}
if we use the coefficients $C,B$ and $A$
in Table \ref{table:ABC}.
We also find a similar agreement for the $O(n)+S$ models.

\begin{figure}[tb]
\begin{center}
\includegraphics[width=7cm]{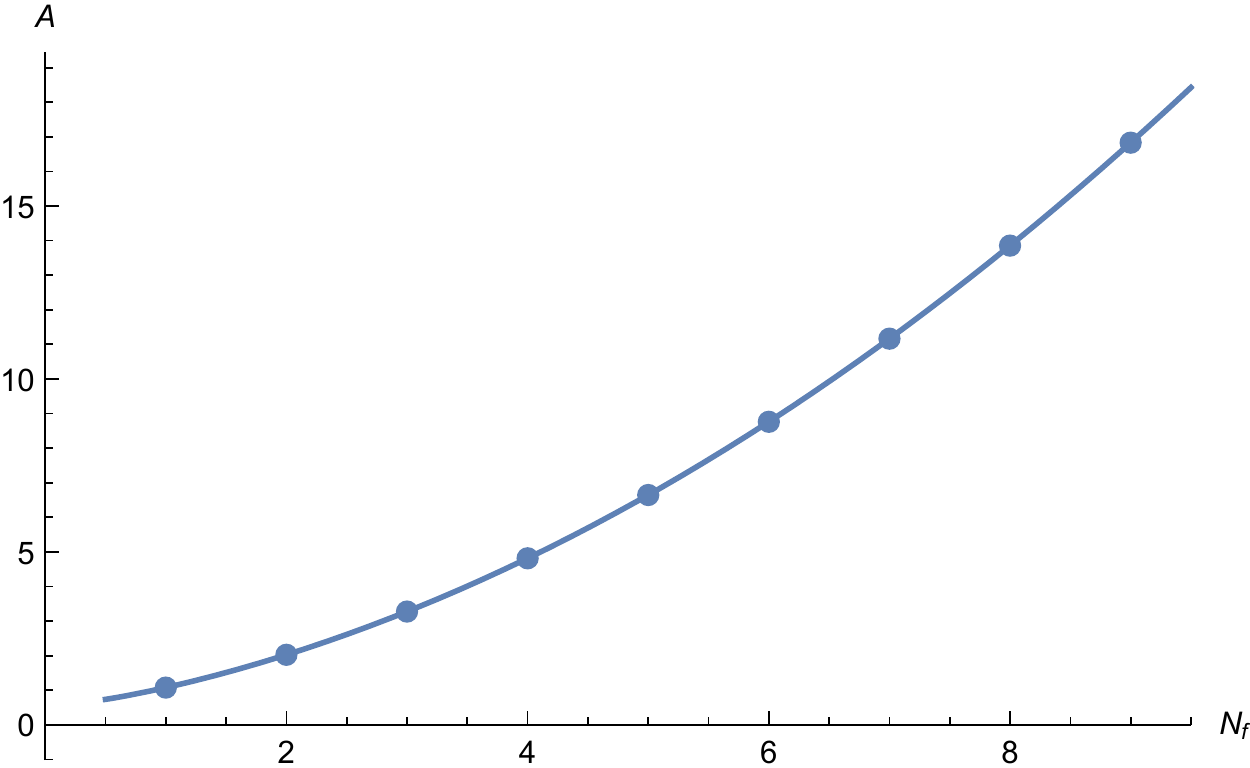}
\end{center}
 \caption{We show the plot of constant $A$
 as a function of $N_f$ for the $O(2N+1)+A$ model.
The dots are the numerical values of $A$ estimated by using
\eqref{Aestimate} for $N_f=1,2,\cdots,9$, while the solid curve
is the plot of $A$ in Table \ref{table:ABC}.
}
\label{fig:Aconst}
\end{figure}

Let us explain in more detail how we found the results in Table
\ref{table:ABC}.
The coefficient $B$ can be found
easily by matching
the Airy function \eqref{Zpert}
with the exact values of $Z(N,N_f)$,
since the $N_f$-dependence of $B$ is relatively simple.
On the other hand, the constant $A$ 
is a complicated function of $N_f$.
To find the constant $A$ as a function of $N_f$,
first
we estimated
the numerical values of  $A$ by
\begin{equation}
 A\approx \log\left[
\frac{Z(N,N_f)}{C^{-\frac{1}{3}}\text{Ai}\Bigl[C^{-\frac{1}{3}}(N-B)\Bigr]}\right],\quad(N\gg1),
\label{Aestimate}
\end{equation}
for $N$ as large as possible.
In practice, we set $N=N_\text{max}$ in \eqref{Aestimate}
where $N_\text{max}$ is the maximal value of $N$ that
the exact values of $Z(N,N_f)$ is available.
In this way, we obtained the constant $A$ for various values of $N_f$'s.
Then, assuming that $A$ is written as a linear combination of
$A_c(k)$ with some $k$'s,
we fixed the coefficients of this linear combination\footnote{Note that $\log2$ in Table \ref{table:ABC} is also written as
 a linear combination of $A_c(2)$ and $A_c(4)$, as shown in \eqref{Ac24}.}, 
and finally we arrived at the expressions of $A$ in Table \ref{table:ABC}.
In Figure \ref{fig:Aconst}, we show the plot of constant $A$ for the $O(2N+1)+A$ model
as an example.
We can clearly see a nice agreement between
the numerical values of $A$ at integer $N_f$ estimated by using \eqref{Aestimate}
and our proposal of $A$ in Table \ref{table:ABC}.
We also find a similar agreement for the other models in Table \ref{table:ABC}.

In section \ref{sec:WKB1},
we will derive the coefficients $B$ and $A$ 
of the $O(2N+1)+A$ model from the WKB expansion.
For other models,
we could not find 
a systematic method to compute $B$ and $A$.

From Table \ref{table:ABC}, we find
the following interesting relations between the $O(2N)$ models and the $O(2N+1)$ models
\begin{equation}
\begin{aligned}
B^{O(2N)+A}-B^{O(2N+1)+A}&=\hf,\\
B^{O(2N)+S}-B^{O(2N+1)+S}&=\hf,\\
A^{O(2N)+A}-A^{O(2N+1)+A}&=-(N_f+1)\log2,\\
A^{O(2N)+S}-A^{O(2N+1)+S}&=-(N_f+2)\log2.
\end{aligned}
\label{diff-AB}
\end{equation}
In section \ref{sec:WKB2}, we will argue that
the right hand side of these relations can be naturally interpreted
as the contributions of orientifold plane.

\section{Non-perturbative corrections}
\label{sec:non-pert}

In this section, we study
the non-perturbative part $J_\text{np}(\mu)$
of grand potential.
To find the coefficients in $J_\text{np}(\mu,N_f)$ 
we follow the procedure in \cite{HMO2}.
First we expand $e^{J_\text{np}(\mu)}$ as
\begin{equation}
 e^{J_\text{np}(\mu)}=1+\sum_{w>0,n}g_{w,n}\mu^n e^{-w\mu},
\label{1+sum}
\end{equation}
where $g_{w,n}$ are some
$\mu$-independent coefficients.
Then the non-perturbative part $Z_\text{np}(N,N_f)$
of partition function is written as a sum of Airy functions and their derivatives
\begin{equation}
\begin{aligned}
 Z_\text{np}(N,N_f)&=Z(N,N_f)-Z_\text{pert}(N,N_f)=\int_{\mathcal{C}}\frac{d\mu}{2\pi i}
e^{J_\text{pert}(\mu)-N\mu}(e^{J_\text{np}(\mu)}-1)\\
&=C^{-\frac{1}{3}}e^A\sum_{w>0,n}g_{w,n}  (-\del_N)^n\text{Ai}\Bigl[C^{-\frac{1}{3}}(N+w-B)\Bigr]. 
\end{aligned}
\end{equation}
By matching the exact values of $Z(N,N_f)$
with the above expansion of $Z_\text{np}(N,N_f)$,
we can fix the coefficients
of $J_\text{np}(\mu)$
order by order in the 
weight $w$ of instantons.
Using this method, we find the non-perturbative corrections
for various $N_f$'s, which are summarized in
Appendix \ref{se:Jnp}.

\subsection{$O(n)+A$}
Let us first consider the model $O(n)+A$ where $n=2N$ or $n=2N+1$.
From the result in Appendix \ref{se:O-even+A} and \ref{se:O-odd+A}, we conjecture that there are three
types of instantons
\begin{equation}
 \mathcal{O}(e^{-\frac{2\mu}{N_f}}),\quad
\mathcal{O}(e^{-2\mu}),\quad
\mathcal{O}(e^{-\mu}).
\label{O+A-type}
\end{equation}
The first two types have natural analogues in the $N_f$ matrix model
\cite{GM,Hatsuda:2014vsa}.
On the other hand, the last one in \eqref{O+A-type}
has no counterpart in the $N_f$ matrix model, hence
it is natural to interpret it as the effect of orientifold plane.
Following \cite{GM,Hatsuda:2014vsa}, let us call
the first two types in \eqref{O+A-type}
worldsheet instantons and membrane instantons, respectively.
For the last type in \eqref{O+A-type},  we will call  
them ``half instantons''.
Note that the weight of worldsheet instanton
in the $N_f$ matrix model is $e^{-4\mu/N_f}$,
which is related to the worldsheet instanton in our case \eqref{O+A-type}
by a rescaling $N_f\to 2N_f$.

\paragraph{Worldsheet instanton}
We conjecture that the worldsheet instanton
corrections are given by
\begin{equation}
 \begin{aligned}
 J_\text{WS}(\mu,N_f)&=-\frac{N_f+2}{2\pi\sin\frac{\pi}{N_f}}\left(\frac{2\mu}{N_f}+1\right)e^{-\frac{2\mu}{N_f}}\\
+&\lf[
-\frac{(N_f+2)^2}{2\pi^2}\left(\frac{2\mu}{N_f}+1\right)^2
+\frac{4(N_f+2)^2-N_f^2}{8\pi N_f}\frac{\sin\frac{3\pi}{N_f}}{ \sin\frac{\pi}{N_f}\sin\frac{2\pi}{N_f}}
\left(\frac{4\mu}{N_f}+1\right)\ri.\\
&\lf.\qquad
-\frac{(N_f+2)^2}{2N_f^2}\frac{\sin\frac{6\pi}{N_f}}{\sin\frac{\pi}{N_f}\sin\frac{2\pi}{N_f}\sin\frac{3\pi}{N_f}}\right]e^{-\frac{4\mu}{N_f}}  
+\mathcal{O}(e^{-\frac{6\mu}{N_f}}),
 \end{aligned}
\label{WS-O+A}
\end{equation}
for both $O(2N)+A$ and $O(2N+1)+A$ models.
For instance, for $N_f=6$
one can see that \eqref{WS-O+A}
correctly reproduces the result of $J_\text{np}(\mu,6)$
in \eqref{O-even-A-intNf} and \eqref{O-odd-A-intNf}.
Note that \eqref{WS-O+A}
is very similar to the worldsheet instantons in the $N_f$ matrix model
\cite{Hatsuda:2014vsa}
and those in the $(1,q)$-model at $k=2$ \cite{Hatsuda:2015lpa}.

\begin{figure}[htb]
\begin{center}
\begin{tabular}{cc}
\includegraphics[width=7cm]{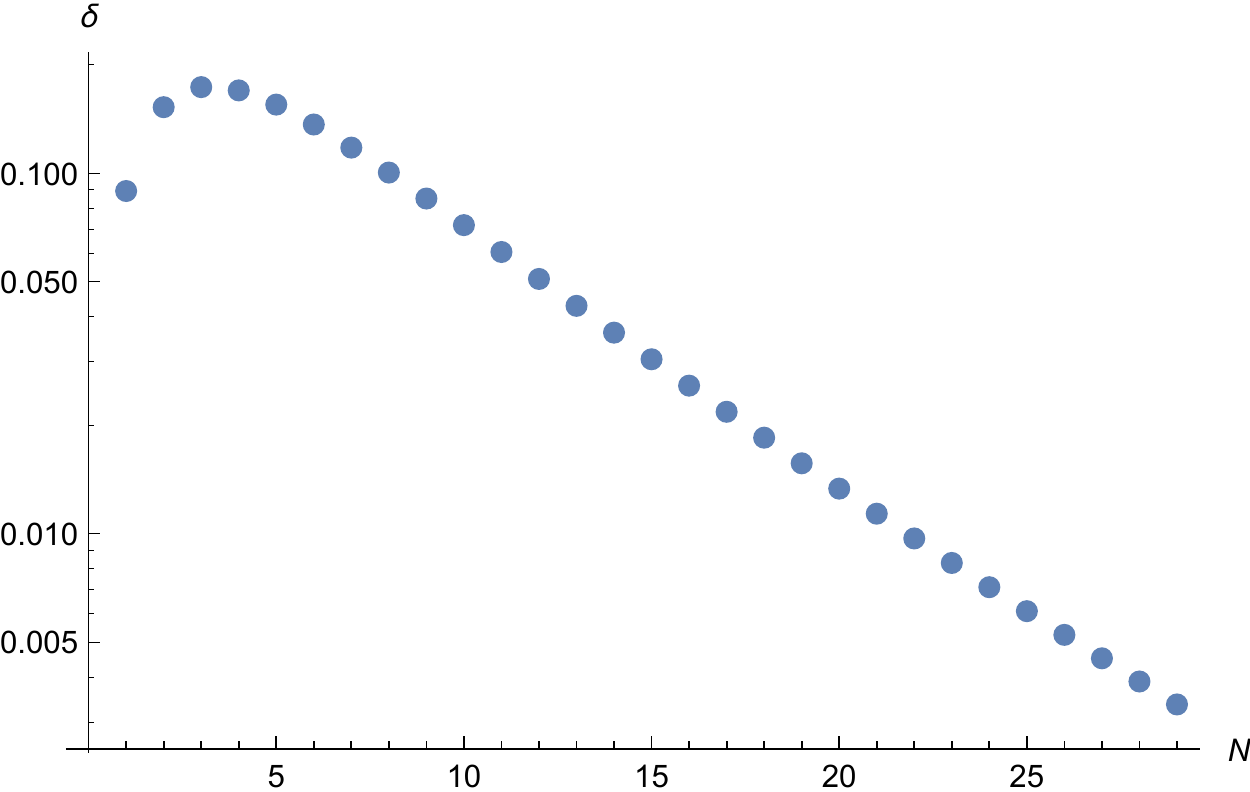}
\hspace{10mm}
&
\includegraphics[width=7cm]{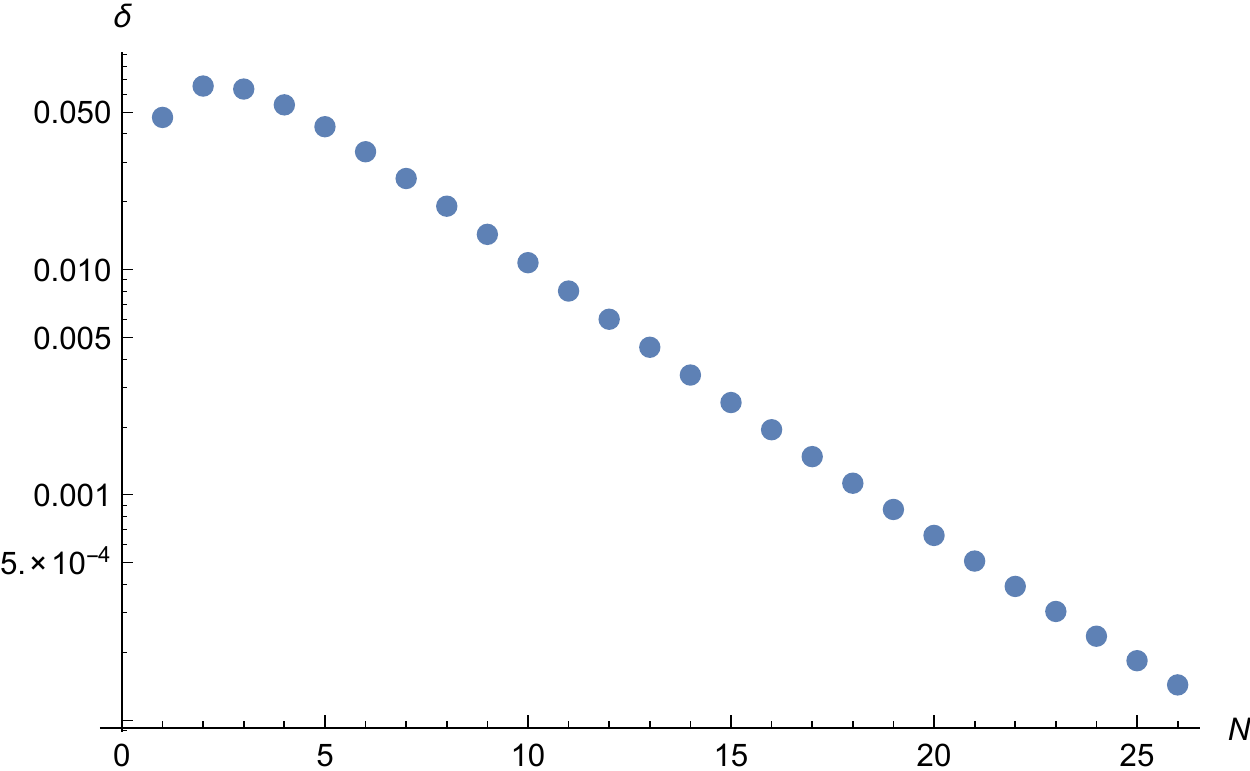}\\
 $N_f=5$
\hspace{10mm}
&
 $N_f=7$\\
\includegraphics[width=7cm]{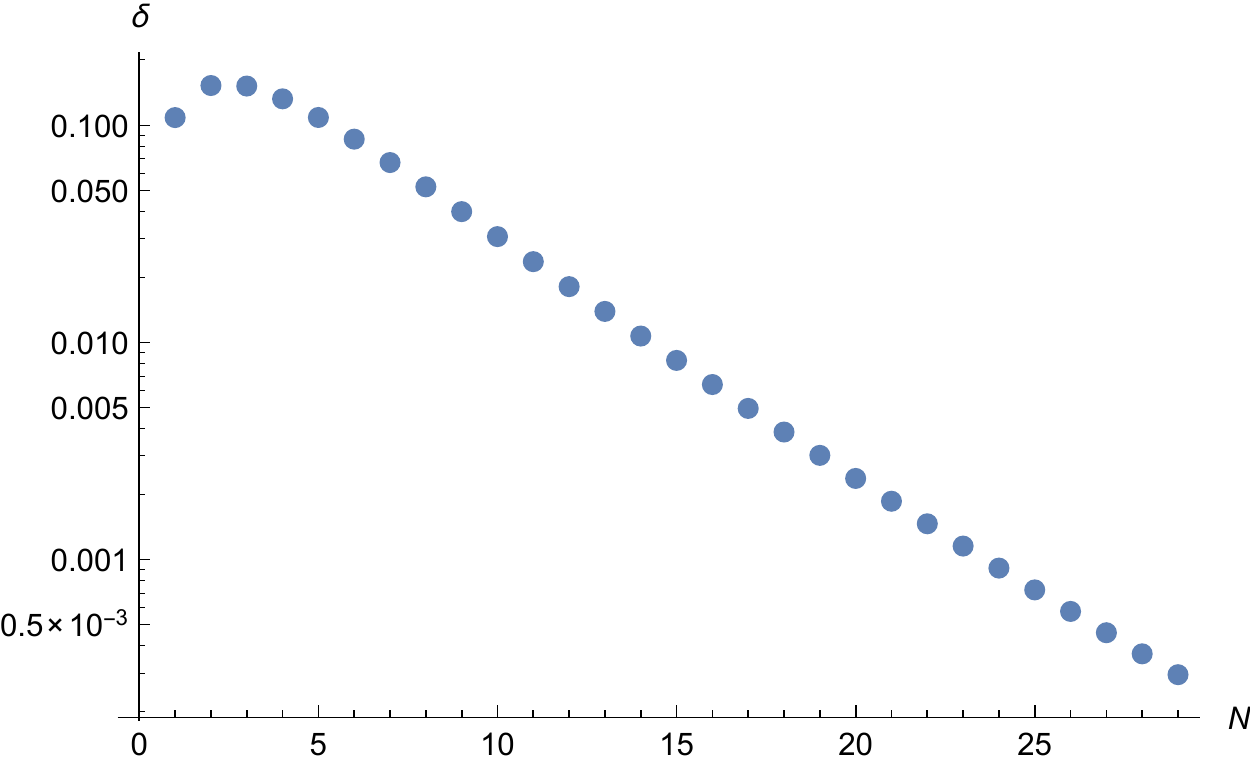}
\hspace{10mm}
&
\includegraphics[width=7cm]{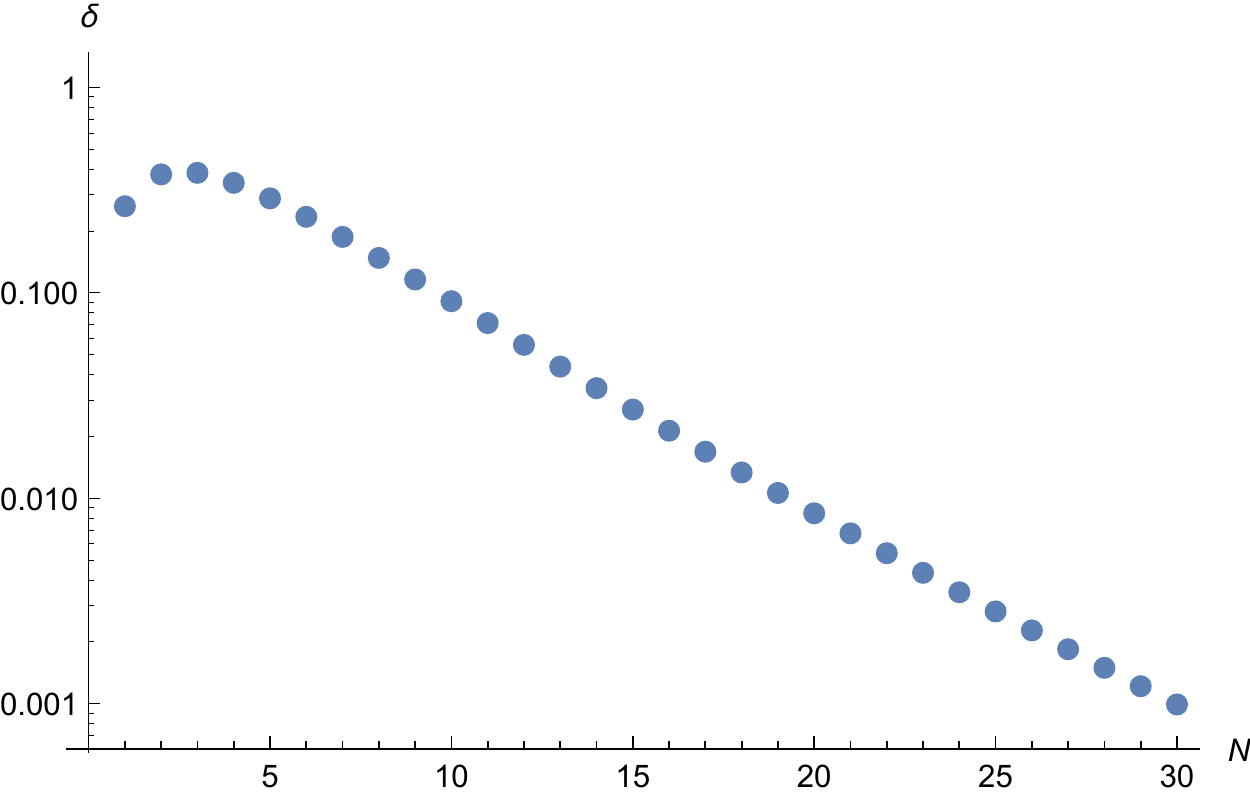}\\
 $N_f=8$
\hspace{10mm}
&
  $N_f=9$
\end{tabular}
\end{center}
  \caption{We show  the plot of quantity $\delta$
 in \eqref{delta-def} for $N_f=5,7,8,9$ in the $O(2N+1)+A$ model.
Note that the vertical axis is log scale.
}
  \label{fig:delta}
\end{figure}

We can check this conjecture \eqref{WS-O+A} in the same way as 
in \cite{Hatsuda:2014vsa}.
We first notice that
for $N_f>4$ the worldsheet 2-instanton 
factor $e^{-4\mu/N_f}$ is larger than the factor $e^{-\mu}$
of half instanton in \eqref{O+A-type}.
Thus,
the non-perturbative part
of canonical partition function
has the following expansion
\begin{equation}
 Z_\text{np}=Z_\text{WS}^{(1)}+Z_\text{WS}^{(2)}+(\text{subleading~corrections}),\quad(N_f>4),
\end{equation} 
where $Z_\text{WS}^{(1)}$
and $Z_\text{WS}^{(2)}$
are the contributions of worldsheet 1-instanton and 
worldsheet 2-instanton to the canonical partition
function.
Now let us consider the following quantity
\begin{equation}
 \delta=\frac{Z-Z_\text{pert}-Z_\text{WS}^{(1)}-Z_\text{WS}^{(2)}}{Z_\text{pert}}
e^{\frac{4\mu_*}{N_f}},
\label{delta-def}
\end{equation}
where $\mu_*$ is given by \eqref{mu-star}. 
If our conjecture of worldsheet instantons \eqref{WS-O+A}
is correct, $\delta$
should be exponentially small in the large $N$
limit.
In Figure \ref{fig:delta}, we plot the quantity $\delta$
for $N_f=5,7,8,9$
in the $O(2N+1)+A$ model.
As we can see in  Figure \ref{fig:delta},
$\delta$
indeed decays exponentially as $N$ becomes large.
We have also checked this behavior for the $O(2N)+A$ model.
We should  also mention that we
have performed a similar
checks for other types of instantons
studied below.

\paragraph{Half instanton}
From the results in Appendix \ref{se:O-even+A} and \ref{se:O-odd+A},
we conjecture that the half instantons are given by
\begin{equation}
 \begin{aligned}
   J_{\text{half}}(\mu,N_f)&=\varepsilon \frac{(\mu+1)}{\pi}2^{N_f}e^{-\mu}+\left[-\frac{(\mu+1)^2}{\pi^2}
+\frac{N_f}{4}\right]2^{2N_f}e^{-2\mu}\\
&\quad+\varepsilon \left[\frac{4(\mu+1)^2}{3\pi^3}-\frac{N_f(\mu+1)}{\pi}\right]
2^{3N_f}e^{-3\mu}+\mathcal{O}(e^{-4\mu}),
 \end{aligned}
 \label{half-inst}
\end{equation}
where $\varepsilon$
is a sign depending on the parity of $n$ of the gauge group $O(n)$
\begin{equation}
 \varepsilon=(-1)^n.
\label{signep}
\end{equation}
Let us take a closer look at these corrections.
For instance, for the $N_f=2$ case,
the order $\mathcal{O}(e^{-\mu})$ term 
comes from the 1-worldsheet instanton $J_\text{WS}^{(1)}$
and the 1-half instanton $J_\text{half}^{(1)}$,
and our conjecture of worldsheet instantons \eqref{WS-O+A} and half instantons
\eqref{half-inst}
correctly reproduce the results in \eqref{O-even-A-intNf}
and \eqref{O-odd-A-intNf}
\begin{equation}
J_\text{WS}^{(1)}(\mu,2)+J_\text{half}^{(1)}(\mu,2)
=\frac{(4\varepsilon-2)(\mu+1)}{\pi}e^{-\mu}.
\end{equation}
Also, for the $N_f=4$ case,
the order $\mathcal{O}(e^{-\mu})$ term comes from
the 2-worldsheet instanton $J_\text{WS}^{(2)}$
and the 1-half instanton $J_\text{half}^{(1)}$,
and the sum of these two contributions
correctly reproduces the results in \eqref{O-even-A-intNf}
and \eqref{O-odd-A-intNf}
\begin{equation}
J_\text{WS}^{(2)}(\mu,4)+J_\text{half}^{(1)}(\mu,4)
=\left[-\frac{9(\mu+2)^2}{2\pi^2}+\frac{(4+16\varepsilon)(\mu+1)}{\pi}+\frac{9}{4}\right]e^{-\mu}.
\end{equation}
We should stress that our conjecture of half instantons \eqref{half-inst}
is consistent with all results in Appendix \ref{se:O-even+A} and \ref{se:O-odd+A}
in a very non-trivial way.

\paragraph{Bound states}
The results in Appendix \ref{se:O-even+A} and \ref{se:O-odd+A}
suggest that there are various types of bound state contributions.
The existence of such bound state contributions are first observed in
ABJM theory \cite{HMO2,HMO-bound}.
Let us denote the
bound state of $\ell$-worldsheet instanton,
$m$-membrane instantons, and $n$-half
instantons as $J^{(\ell,m,n)}(\mu,N_f)$. Namely,
\begin{equation}
J^{(\ell,m,n)}(\mu,N_f)\propto
 e^{-\frac{2\ell\mu}{N_f}-2m\mu-n\mu}.
\end{equation}

For instance, there seems to exist a bound state 
of 1-worldsheet instanton $e^{-2\mu/N_f}$
and 1-half instanton $e^{-\mu}$.
From the coefficient of $\mathcal{O}(e^{-2\mu/N_f-\mu})$
term
for $N_f=3,1/2,3/2,5/2$ in Appendix \ref{se:O-even+A} and \ref{se:O-odd+A},
we conjecture
\begin{align}
 J^{(1,0,1)}(\mu,N_f)=\varepsilon\frac{2^{N_f+1}}{\sin\frac{\pi}{N_f}}e^{-\frac{2\mu}{N_f}-\mu}.
 \label{bound}
\end{align} 

There should also exist
a bound state $J^{(0,1,1)}$
of 1-membrane instanton $e^{-2\mu}$ and 1-half instanton $e^{-\mu}$
in order to reproduce the finite term of $N_f=1$
at order $\mathcal{O}(e^{-3\mu})$
\begin{equation}
 \lim_{N_f\to1}\Big[J^{(1,0,1)}(\mu,N_f)+
J^{(0,1,1)}(\mu,N_f)
+J^{(0,0,3)}(\mu,N_f)\Big]=
\varepsilon \frac{88\mu+52/3}{3\pi}e^{-3\mu}.
\label{Nf1lim}
\end{equation}
Note that $J^{(1,0,1)}$ in \eqref{bound}
has a pole at $N_f=1$, while
the 3-half instanton $J^{(0,0,3)}$
in \eqref{half-inst} is regular at $N_f=1$.
For the equation \eqref{Nf1lim}
to make sense,
the bound state of membrane instanton and half instanton
$J^{(0,1,1)}$ should cancel the pole coming
from $J^{(1,0,1)}$ in \eqref{bound}.
This {\it pole cancellation mechanism},
first discovered in the ABJM theory \cite{HMO2},
gives a constraint for the possible form of the coefficient
of $J^{(0,1,1)}$. But this condition alone is not strong enough to determine $J^{(0,1,1)}$.

\paragraph{Membrane instanton}
We do not have a direct information
of the coefficient of membrane
instantons in
the results of Appendix \ref{se:O-even+A} and \ref{se:O-odd+A}.
However,
from the pole cancellation mechanism
and the information of finite terms
at $N_f=1,2$, we can make a conjecture of membrane 1-instanton, as we will see below.

From the expression of the membrane instanton in the $N_f$ matrix model \cite{Hatsuda:2015lpa},
it is natural to conjecture that
the coefficient of 1-membrane
instanton is
proportional to $2\mu+1$
\begin{align}
 J_{\text{M2}}(\mu,N_f)&=b_1(N_f)(2\mu+1)e^{-2\mu}+\mathcal{O}(e^{-4\mu}).
\end{align}
From \eqref{O-even-A-halfNf} and \eqref{O-odd-A-halfNf},
we observe that the 1-membrane instanton term
is absent for half-integer $N_f$.
Thus, the coefficients $b_1(N_f)$
should vanish at half-integer $N_f$
\begin{equation}
 b_1(N_f)=0,\quad\Bigl(N_f\in\mathbb{Z}_{\geq0}+\hf\Bigr).
\end{equation}
Also, 
the result of $N_f$ matrix model in
\cite{Hatsuda:2014vsa}
suggests that
$b_1(N_f)$
is given by a certain combination of gamma-functions.
Furthermore, $J_\text{M2}^{(1)}$ should reproduce the finite terms at $N_f=1,2$
in \eqref{O-even-A-intNf} and \eqref{O-odd-A-intNf}
\begin{equation}
 \begin{aligned}
 \lim_{N_f\to1}\Bigl[J_\text{WS}^{(1)}+J_{\text{half}}^{(2)}
 +J_{\text{M2}}^{(1)}\Bigr]&=
 \left[-\frac{10\mu^2+7\mu+7/2}{\pi^2}+1\right]e^{-2\mu},\\
 \lim_{N_f\to2}\Bigl[J_\text{WS}^{(2)}+J_{\text{half}}^{(2)}
 +J_{\text{M2}}^{(1)}+J^{(1,0,1)}\Bigr]&=
 \left[-\frac{39\mu^2+63\mu/2+63/4}{\pi^2}+14+8\varepsilon\right]e^{-2\mu}. 
 \end{aligned}
\label{finite}
\end{equation}
Since $J_\text{WS}^{(1)}$ and $J_\text{WS}^{(2)}$ have poles at $N_f=1$ and $N_f=2$,
respectively, 1-membrane instanton $J_\text{M2}^{(1)}$
should cancel those poles and give the finite terms in the right hand side of \eqref{finite}.
From this pole cancellation
condition and other
conditions mentioned above,
we conjecture that $J_\text{M2}^{(1)}$
is given by
\begin{align}
 J_\text{M2}^{(1)}(\mu,N_f)=\frac{\Gamma(-N_f)^2}{8\pi^2\Gamma(-2N_f-1)}(2\mu+1)e^{-2\mu}.
\label{M2-cand}
\end{align}
One can show that our conjecture \eqref{M2-cand}
indeed reproduces the right hand side of \eqref{finite} and vanishes
when $N_f$ is positive half-integer, as required.
It would be nice to see if
our conjecture \eqref{M2-cand} of 1-membrane instanton is correct or not, by computing it from 
the WKB expansion as in \cite{Hatsuda:2015lpa,Hatsuda:2014vsa}.

\subsection{$O(n)+S$}
Next consider the model $O(n)+S$ with $n=2N$ or $n=2N+1$.
From the results in Appendix \ref{se:O-even+S} and \ref{se:O-odd+S}, 
there seems to be two types of instantons
\begin{equation}
 \mathcal{O}(e^{-\frac{2\mu}{N_f}}),\quad
\mathcal{O}(e^{-2\mu}),
\end{equation}
which are the analogue of worldsheet instantons and membrane instantons
in the $N_f$ matrix model.
There are no $\mathcal{O}(e^{-\mu})$ term in this case.

We conjecture that the
worldsheet instanton corrections are given by
\begin{equation}
 \begin{aligned}
   J_\text{WS}(\mu,N_f)=&-\frac{N_f}{2\pi\sin\frac{\pi}{N_f+2}}\left(\frac{2\mu}{N_f+2}+1\right)e^{-\frac{2\mu}{N_f+2}}\\
&+\left[-\frac{N_f^2}{2\pi^2}\left(\frac{2\mu}{N_f+2}+1\right)^2
+\frac{4N_f^2-(N_f+2)^2}{8\pi (N_f+2)}\frac{\sin\frac{3\pi}{N_f+2}}{ \sin\frac{\pi}{N_f+2}\sin\frac{2\pi}{N_f+2}}
\left(\frac{4\mu}{N_f+2}+1\right)\ri.\\
&\lf.\qquad
-\frac{N_f^2}{2(N_f+2)^2}\frac{\sin\frac{6\pi}{N_f+2}}{\sin\frac{\pi}{N_f+2}\sin\frac{2\pi}{N_f+2}\sin\frac{3\pi}{N_f+2}}\right]e^{-\frac{4\mu}{N_f+2}}
+\mathcal{O}(e^{-\frac{6\mu}{N_f+2}}).
 \end{aligned}
\label{WS-O+S}
\end{equation}
We have performed a similar check as in Figure \ref{fig:delta}
for our conjecture \eqref{WS-O+S},
and confirmed that \eqref{WS-O+S} correctly
reproduce the large $N$ behavior of exact values $Z(N,N_f)$
 for various $N_f$'s.

As for the membrane instantons, we were unable to determine their coefficient
from the data in Appendix \ref{se:O-even+S} and \ref{se:O-odd+S} alone.
It would be interesting to study the structure of instanton
corrections in this model
further.

\section{WKB expansion (I)}
\label{sec:WKB1}
As discussed in \cite{Assel:2015hsa},
we can compute the
coefficients
$C$ and $B$ in the perturbative
part of grand potential by {\it formally}
introducing the Planck constant
$\hbar$ and performing the small $\hbar$ expansion,
although the physical theory corresponds to
$\hbar=2\pi$.
Since the coefficients $C$ and $B$
receive corrections only up to order $\mathcal{O}(\hbar^0)$
and $\mathcal{O}(\hbar^2)$, respectively,
we can fix $C$ and $B$
by computing the first two terms of WKB expansion
and simply setting $\hbar=2\pi$ at the end.
On the other hand,
the coefficient $A$
receives all order corrections in $\hbar$,
hence it is not obvious if we can find
the constant $A$ in $J_\text{pert}(\mu)$
from this formal WKB expansion.
Nevertheless,
as we will see below,
at least for the $O(2N+1)+A$ model
we can guess the all order
expression of the WKB expansion of $A$,
and by setting
$\hbar=2\pi$ we find the constant
$A$ in a closed form.
For models other than $O(2N+1)+A$,
we found difficulty in computing the leading
(classical) term in the WKB expansion.
Therefore, in this section we will focus on the $O(2N+1)+A$ model.
Note that, the $O(2N+1)+A$ model is related to the $USp(2N)+A$ model
by a shift of $N_f$ \eqref{O-Sp-relation}, which 
in turn is 
dual to a $\h{D}$-type quiver theory by mirror symmetry \cite{deBoer:1996mp}. 

\subsection{WKB expansion in $O(2N+1)+A$ model}
As discussed in \cite{Assel:2015hsa},
the density matrix $\rho(x,y)$ in
\eqref{rhoxy} for the $O(2N+1)+A$ model
can be written as a matrix element $\bra x|\rho|y\ket$
of the quantum mechanical operator
$\rho=e^{-H}$
of the following form\footnote{The relation between the
density matrix in \eqref{rhoxy} with $a=c=1,b=d=0$,
and the operator in \eqref{rho-D}
can be shown by using
the relation \cite{Assel:2015hsa}
\begin{equation}
 \left(\sinh\frac{\h{x}}{2} \frac{1}{\cosh\frac{\h{p}}{2}}\cosh\frac{\h{x}}{2}+
  \cosh\frac{\h{x}}{2} \frac{1}{\cosh\frac{\h{p}}{2}}\sinh\frac{\h{x}}{2}\right)\frac{1+R}{2}
 =\sinh\frac{\h{x}}{2}\frac{1+R}{2\cosh\frac{\h{p}}{2}}
 \frac{\cosh\h{x}}{\cosh\frac{\h{x}}{2}}.
\end{equation}
} 
\begin{equation}
\begin{aligned}
 \rho&=\rho_D\frac{1+R}{2},\nn
\rho_D&=\frac{2\sinh\frac{\h{x}}{2}}{(2\cosh\frac{\h{x}}{2})^{2N_f+3}}
\left(\sinh\frac{\h{x}}{2} \frac{1}{\cosh\frac{\h{p}}{2}}\cosh\frac{\h{x}}{2}+
\cosh\frac{\h{x}}{2} \frac{1}{\cosh\frac{\h{p}}{2}}\sinh\frac{\h{x}}{2}\right), 
\end{aligned}
\label{rho-D}
\end{equation}
where $\h{x}$ and $\h{p}$ are the canonical variables
obeying
\begin{equation}
 [\h{x},\h{p}]=i\hbar,\qquad \hbar=2\pi,
\end{equation}
and $R$ in \eqref{rho-D} is the reflection operator
flipping the sign of $x$
\begin{equation}
 R|x\ket=|-x\ket.
\end{equation}
Note that in \cite{Mezei:2013gqa}
a different expression of
operator $\rho$ was used.
We will consider the operator in \cite{Mezei:2013gqa}
in the next section.
One advantage of $\rho_D$ in \eqref{rho-D}
is that it has not only
a reflection symmetry $[\rho_D,R]=0$,
but also has a property that
its trace
with $R$ insertion vanishes
\cite{Assel:2015hsa}
\begin{equation}
  \Tr(\rho_D^\ell R)=0,\qquad (\ell=1,2,\cdots).
\end{equation}
This implies that
 the grand partition function is written as
 \begin{equation}
  \Xi(\mu)=\text{Det}(1+e^\mu\rho)=
   \sqrt{\text{Det}(1+e^\mu\rho_D)},
 \end{equation}
 and the grand potential is given by
 \begin{equation}
  J(\mu)=\hf\sum_{\ell=1}^\infty \frac{(-1)^{\ell-1}e^{\ell\mu}}{\ell}
   \Tr\rho_D^\ell.
\label{JinTr}
 \end{equation}

As noticed in \cite{Hatsuda:2015oaa},
the WKB expansion of the grand potential
\eqref{JinTr}
is most easily obtained
from the WKB expansion of the {\it spectral trace}
$\Tr\rho_D^s$.
By analytically continuing $\Tr\rho_D^s$
from integer $s$ to arbitrary complex $s$,
the grand potential
is written as a Mellin-Barnes type integral
\begin{equation}
 J(\mu)=-\hf\int_\ga\frac {ds}{2\pi i}\Ga(-s)\Ga(s)e^{s\mu}\Tr\rho_D^s,
\label{JMellin}
\end{equation}
where the integration contour $\ga$ is parallel to the imaginary
axis with $0<\Re(s)<1$.
By picking up poles at positive integers $s=\ell\in\mathbb{Z}_{>0}$,
we recover
\eqref{JinTr}.
On the other hand,
deforming the contour in the direction $\Re(s)\leq0$,
we can find a large $\mu$ expansion of $J(\mu)$.
The WKB expansion of spectral trace takes the following form
\begin{equation}
\begin{aligned}
 \Tr\rho_D^s&=Z_0(s)D(s,\hbar),\\
D(s,\hbar)&=1+\sum_{n=1}^\infty D_n(s)\hbar^{2n}.
\end{aligned}
\label{TrWKB}
\end{equation}
The leading term 
$Z_0(s)$ is given by
the classical phase space integral,
simply replacing the operators $(\h{x},\h{p})$ in \eqref{rho-D}
by classical commuting variables $(X,P)$
\begin{align}
Z_0(s)=\int\frac{dXdP}{2\pi\hbar} \left[\frac{(2\sinh\frac{X}{2})^2}{2\cosh\frac{P}{2}(2\cosh\frac{X}{2})^{2N_f+2}}\right]^{s}=\frac{1}{2\pi\hbar}\frac{2\Ga(s/2)^2\Ga(2s)\Ga(N_fs)\Ga((N_f+1)s)}{\Ga(s)^2\Ga(2(N_f+1)s)}.
\label{Z0}
\end{align}
From the WKB expansion of the spectral trace \eqref{TrWKB},
we can easily find the WKB expansion
of grand potential by replacing $s$ 
in $D(s,\hbar)$ 
by the $\mu$-derivative $\del_\mu$,
and acting it on the leading term
\begin{equation}
 J(\mu)=D(\del_\mu,\hbar)
J_0(\mu),
\label{doponJ}
\end{equation}
where $J_0(\mu)$ is the leading  term in the WKB expansion
of grand potential
\begin{equation}
 J_0(\mu)=-\hf\int_{\ga}\frac {ds}{2\pi i}\Ga(-s)\Ga(s)e^{s\mu}Z_0(s).
\end{equation}

Now, let us move on to the computation of $D_n(s)$ in \eqref{TrWKB}.
In many examples of $d=3~\mathcal{N}=4$ theories
\cite{Moriyama:2014waa,Hatsuda:2015lpa}, it turned out that $D_n(s)$ was a rational function of $s$. 
Therefore, 
it is natural to assume that this is also the case for our expansion
\eqref{TrWKB}.
Then, the easiest way to determine $D_n(s)$ is
to make an ansatz that $D_n(s)$ is a rational function of $s$,
and fix the coefficients in the ansatz
by matching the WKB expansion of $\Tr\rho_D^\ell$ for integer $\ell$
from $\ell=1$ to some $\ell=\ell_\text{max}$,
where we choose $\ell_\text{max}$ as the number of independent coefficients
in the ansatz of $D_n(s)$.
Once we determined $D_n(s)$ in this way,
we can check the agreement of $\Tr\rho_D^\ell$ and $D_n(\ell)$
 for $\ell>\ell_\text{max}$.

 To compute the WKB expansion of $\Tr\rho_D^\ell$
for integer $\ell$,
 it is convenient
 to use the Wigner transform of the operator $\rho_D$.
 In general, the Wigner transform $O_W$  is defined by
\begin{align}
 O_W=\int dy e^{\frac{iPy}{\hbar}}\Big\bra X-\hf y\Big|O\Big|X+\hf y\Big\ket.
 \label{Wignerdef}
\end{align} 
As explained in Appendix \ref{se:Wigner},
the Wigner transform of $\rho_D$ is given by
\begin{equation}
 (\rho_D)_W=\frac{2\sinh\frac{X}{2}}{(2\cosh\frac{X}{2})^{2N_f+3}}\star
\frac{\sinh X}{\cosh\frac{P}{2}}.
\label{rhoDW}
\end{equation}
Using the property of Wigner transformation
\begin{equation}
 (AB)_W=A_W\star B_W \equiv A_W e^{\frac{i\hbar}{2}(\overleftarrow{\del_X}\overrightarrow{\del_P}
-\overleftarrow{\del_P}\overrightarrow{\del_X})} B_W,
  \label{Wprod}
\end{equation}
the Wigner transform of the $\ell^\text{th}$
power of $\rho_D$ is given by the star-product of
$(\rho_D)_W$'s
\begin{equation}
 (\rho_D^\ell)_W=\underbrace{ (\rho_D)_W \star\cdots\star (\rho_D)_W }_\ell.
\end{equation}
Now, let us consider the WKB expansion of  $(\rho_D^\ell)_W$
\begin{equation}
 (\rho_D^{\ell})_W=\sum_{m=0}^\infty \rho^{\ell}_{(m)}\hbar^{m}.
\label{rhoWexp}
\end{equation}
From the obvious relation
$(\rho_D^{\ell})_W=(\rho_D)_W\star(\rho_D^{\ell-1})_W$,
the coefficient $\rho^\ell_{(m)}$
of this expansion can be computed recursively in $\ell$
\begin{equation}
 \begin{aligned}
   \rho^{\ell}_{(m)}&=\sum_{n_1+n_2+n_3=m}
  \rho_{(n_1)}\frac{(i/2)^{n_2}}{n_2!}
  \Big(\overleftarrow{\del_X}\overrightarrow{\del_P}
-\overleftarrow{\del_P}\overrightarrow{\del_X}\Big)^{n_2} \rho^{\ell-1}_{(n_3)}.
 \end{aligned}
\end{equation}
Using the fact that the trace $\Tr\rho_D^\ell$
is written as a classical phase space integral
of $(\rho_D^\ell)_W$
\begin{equation}
 \Tr\rho_D^\ell=\int\frac{dXdP}{2\pi\hbar}
  (\rho_D^\ell)_W,
\label{trint}
\end{equation}
and plugging the WKB expansion of $(\rho_D^{\ell})_W$
\eqref{rhoWexp}
into \eqref{trint}, finally we find the WKB expansion
of the trace $\Tr\rho_D^\ell$.

Using the above method,
we have computed $D_n(s)$
up to $n=13$.
We find that $D_n(s)$ has the following form
\begin{equation}
 D_n(s)=\frac{p_n(s)}{96^n\prod_{j=1}^n(2(N_f+1)s+2j-1)(s+2j-1)},
\end{equation}
where $p_n(s)$ is a $(4n)^{\text{th}}$ order polynomial of $s$.
The first few terms are given by
\begin{equation}
 \begin{aligned}
  p_1(s)&=N_f s^2(1-s)(3+N_f+2(N_f+1)s),\nn
p_2(s)&=\frac{N_fs^3(1-s)}{10}\left[-28 N_f
   (N_f+1)^2 s^4-4 (N_f+1) \left(25 N_f^2+47 N_f+2\right)s^3\ri.\nn
&\quad +\left(-29 N_f^3-438
   N_f^2-549 N_f-56\right) s^2+\left(-41 N_f^3-178 N_f^2-603
   N_f-250\right) s\nn
&\qquad \lf.-6 \left(4 N_f^3+2 N_f^2+7 N_f+67\right)\right],\nn
p_3(s)&=\frac{N_fs^3(1-s)}{105}
\Bigl[372 N_f^2 (N_f+1)^3 s^8+2 N_f (N_f+1)^2 \left(1829
   N_f^2+3059 N_f+164\right) s^7\nn
&\quad+(N_f+1) \left(10807 N_f^4+48062
   N_f^3+47327 N_f^2+5356 N_f+96\right) s^6\nn
&\quad +\frac{1}{2} \left(17423
   N_f^5+231087 N_f^4+586289 N_f^3+460893 N_f^2+80084
   N_f+2976\right) s^5\nn
&\quad+\left(7477 N_f^5+92925 N_f^4+458458
   N_f^3+596733 N_f^2+187615 N_f+9432\right) s^4\nn
&\quad+\frac{1}{2}
   \left(12125 N_f^5+96285 N_f^4+570131 N_f^3+1605111
   N_f^2+1117364 N_f+97224\right) s^3\nn
&\quad+\left(8596 N_f^5+38502
   N_f^4+39601 N_f^3+376104 N_f^2+838273 N_f+188364\right)
   s^2\nn
&\quad+12 \left(768 N_f^5+1104 N_f^4-812 N_f^3+3112
   N_f^2+31519 N_f+29029\right) s\nn
&\quad
+180 \left(16 N_f^5+2 N_f^2+7  N_f+1015\right)
\Bigr].
 \end{aligned}
\end{equation}
From this we can read off the general structure of $p_n(s)$
for $n\geq2$
\begin{align}
 p_n(s)=N_f(1-s)s^3\sum_{j=0}^{4n-4}s^j g_n^{(j)}(N_f),
\end{align} 
where $g_n^{(j)}(N_f)$ is a $(2n-1)^\text{th}$ order polynomial of $N_f$.
Note that $p_1(s)$ is an exception: $p_1(s)$ has a factor $s^2(1-s)$
while $p_n(s)~(n\geq2)$ has a factor $s^3(1-s)$.
As we will see in the next subsection, this is related to the
difference of the $\hbar$ corrections of $C,B$ and $A$.

\subsection{Perturbative part of $O(2N+1)+A$ from WKB expansion}
By deforming the contour $\ga$
to the left half plane $\Re(s)\leq0$ 
in \eqref{JMellin}, we can find the
large $\mu$ expansion of
the grand potential $J(\mu)$.
It turns out that the perturbative part $J_\text{pert}(\mu)$
comes from the pole 
at $s=0$.
The leading contribution of the WKB expansion
reads
\begin{equation}
\begin{aligned}
 &J_{\text{pert},(0)}(\mu)=
-\hf\oint_{s=0}\frac {ds}{2\pi i}\Ga(-s)\Ga(s)e^{s\mu}Z_0(s)\\
&=\frac{\mu^3}{6N_f\pi^2}
+\left(\frac{1}{8N_f}-\frac{N_f+3}{6}\right)\mu
+\frac{\zeta(3)}{4N_f\pi^2}+\frac{(2N_f^2+7N_f+7N_f)\zeta(3)}{\pi^2}.
\end{aligned}
\label{Jpertlead}
\end{equation}
The $\hbar$-corrections can be computed systematically
by applying the relation \eqref{doponJ} to the perturbative part
\begin{equation}
 J_\text{pert}(\mu)=D(\del_\mu,\hbar)
J_{\text{pert},(0)}(\mu).
\end{equation}
Since $J_{\text{pert},(0)}(\mu)$ is a cubic polynomial in $\mu$,
the derivatives $\del_\mu^m$ with $m\geq4$ 
do not contribute to $J_{\text{pert}}(\mu)$.
By expanding $D_n(\del_\mu)$ up to $\del_\mu^3$, we find
\begin{equation}
 \begin{aligned}
  &D_n(\del_\mu)=\cob_{n,1}\frac{N_f(N_f+3)}{96}\del_\mu^2\\
&+\frac{N_f}{4} \frac{(-1)^nB_{2n}B_{2n-2}}{(2n)!2^{2n}}
\Bigl[(2N_f)^{2n-1}+2(2N_f^2+7N_f+7)+2 \cdot 4^{2n-1}- 2^{2n-1}\Bigr]\del_\mu^3
+\mathcal{O}(\del_\mu^4).
\label{Dsmalls}
 \end{aligned}
\end{equation}
We have checked this behavior up to $n=13$ and we believe
that this is true for all $n$.

From the expansion in \eqref{Dsmalls}, one can easily see that
$C$ and $B$ receive corrections only up to $\mathcal{O}(\hbar^0)$
and $\mathcal{O}(\hbar^2)$, respectively.
Acting the differential operator on the leading term 
$J_{\text{pert},(0)}$ and setting $\hbar=2\pi$, 
finally we arrive at the correct $C$ and $B$
of the $O(2N+1)+A$ model in Table \ref{table:ABC} 
\begin{equation}
 \left(1+\frac{N_f(N_f+3)}{96}\del_\mu^2\hbar^2\right)
J_{\text{pert},(0)}\Big|_{\hbar=2\pi}=\frac{\mu^3}{6N_f\pi^2}+\left(\frac{1}{8N_f}-\frac{N_f+3}{8}\right)\mu+\mathcal{O}(\mu^0).
\end{equation}

We can also determine
the constant $A$
by summing over all
order corrections.
From \eqref{Jpertlead}
and \eqref{Dsmalls},
one can easily see that
the constant $A$ is given by
\begin{equation}
\begin{aligned}
 A&=\frac{\zeta(3)}{4N_f\pi^2}+\frac{(2N_f^2+7N_f+7N_f)\zeta(3)}{\pi^2}\\
&+\sum_{n=1}^{\infty}\frac{N_f}{4} \frac{(-1)^nB_{2n}B_{2n-2}}{(2n)!2^{2n}}
\Bigl[(2N_f)^{2n-1}+2(2N_f^2+7N_f+7)+2 \cdot 4^{2n-1}- 2^{2n-1}\Bigr]\del_\mu^3\hbar^{2n}
\frac{\mu^3}{6N_f\pi^2}\Big|_{\hbar=2\pi}\\
&=\frac{\zeta(3)}{4N_f\pi^2}+\frac{(2N_f^2+7N_f+7N_f)\zeta(3)}{\pi^2}\\
&+\qu\sum_{n=1}^{\infty} \frac{(-1)^nB_{2n}B_{2n-2}\pi^{2n-2}}{(2n)!}
\Bigl[(2N_f)^{2n-1}+2(2N_f^2+7N_f+7)+2 \cdot 4^{2n-1}- 2^{2n-1}\Bigr].
\end{aligned}
 \end{equation}
By comparing this with the small $k$ expansion of $A_c(k)$ \cite{KEK}
\begin{equation}
 A_c(k)=\frac{2\zeta(3)}{\pi^2 k}+\sum_{n=1}^\infty \frac{(-1)^n}{(2n)!}
B_{2n}B_{2n-2}\pi^{2n-2}k^{2n-1},
\end{equation}
we find that $A$ is written as
\begin{equation}
 A=\qu A_c(2N_f)+\frac{2N_f^2+7N_f+7}{2}A_c(1)+\frac{2A_c(4)-A_c(2)}{4}.
\label{AinAc}
\end{equation}
Using the explicit values of $A_c(4)$ and $A_c(2)$
from \eqref{Acintk}
\begin{equation}
 A_c(4)=-\frac{\zeta(3)}{4\pi^2}-\hf\log2,\quad
  A_c(2)=-\frac{\zeta(3)}{2\pi^2},
  \label{Ac24}
\end{equation}
one can see that \eqref{AinAc} correctly reproduces
the constant $A$ of the $O(2N+1)+A$ model
in Table \ref{table:ABC}.
\subsection{Comment on the non-perturbative part of $O(2N+1)+A$}
In principle,
we can also study
the non-perturbative corrections $J_\text{np}(\mu)$
from the WKB analysis in the previous subsection.
$J_\text{np}(\mu)$
comes from the poles
on the negative real axis in the Mellin-Barnes representation
\eqref{JMellin}.
For instance,
the leading term of spectral trace $Z_0(s)$
has poles at $s=-1/2$ and $s=-1/N_f$,
and their contribution to $J_\text{np}(\mu)$
is given by
\begin{equation}
\begin{aligned}
 &J_\text{np}(\mu)=\frac{1}{\hbar}D(\del_\mu,\hbar)\left[-\frac{2^{4+N_f}\pi^{\frac{3}{2}}}{\Ga(\qu)^2
}e^{-\frac{\mu}{2}}
-\frac{4(N_f+2)\pi \Ga(\frac{1}{N_f})}{\sin^2\frac{\pi}{2N_f}\Ga(\frac{1}{2N_f})^2}
e^{-\frac{\mu}{N_f}}+\cdots\right] \\
&=\frac{1}{\hbar}
\left[-\frac{2^{4+N_f}\pi^{\frac{3}{2}}}{\Ga(\qu)^2
}D\Bigl(-\hf,\hbar\Bigr)e^{-\frac{\mu}{2}}
-\frac{4(N_f+2)\pi \Ga(\frac{1}{N_f})}{\sin^2\frac{\pi}{2N_f}\Ga(\frac{1}{2N_f})^2}
D\Bigl(-\frac{1}{N_f},\hbar\Bigr)e^{-\frac{\mu}{N_f}}
+\cdots\right]. 
\end{aligned}
\end{equation}
For these two terms,
we find an all order expression of 
the coefficients.
For the $e^{-\mu/N_f}$ term, we find
\begin{equation}
D\Bigl(-\frac{1}{N_f},\hbar\Bigr)=\frac{\hbar}{4}\cot\Bigl(\frac{\hbar}{4}\Bigr)
\exp\left[\sum_{n=1}^\infty \frac{(-1)^{n-1}B_{2n}}{2n(2n+1)!}
\Biggl(2B_{2n+1}\Bigl(\frac{1}{2N_f}\Bigr)-B_{2n+1}\Bigl(\frac{1}{N_f}\Bigr)\Biggr)(N_f\hbar)^{2n}\right].
\label{Dinst}
\end{equation}
We have checked that this agrees with the WKB expansion up to $n=13$.
For the $e^{-\mu/2}$ term,
we observe that the coefficient $D(-1/2,\hbar)$ is independent of $N_f$\footnote{For a generic $s$, $D(s,\hbar)$ has a non-trivial dependence on $N_f$: $D(s,\hbar)=D(s,\hbar;N_f)$. However, it happens to be the case that $D(s,\hbar)$ is independent of $N_f$ at $s=-1/2$.},
\begin{align}
 D\Bigl(-\frac{1}{2},\hbar\Bigr)=1-\frac{\hbar^2}{64}
-\frac{29\hbar^4}{122880}+\frac{53 \hbar^6}{23592960}+\cdots,
\end{align}
hence it is simply given by setting $N_f=2$ in \eqref{Dinst}.
From \eqref{Dinst}, one can see that
the coefficients of $e^{-\mu/N_f}$
and $e^{-\mu/2}$ both vanish in the limit $\hbar\to2\pi$. This is consistent with
our result in Appendix \ref{se:O-odd+A}
that there are no 
$e^{-\mu/N_f}$  and $e^{-\mu/2}$ terms
in the grand potential of $O(2N+1)+A$ model. 
Note that \eqref{Dinst}
is essentially equal to a combination of $q$-gamma functions
$\Ga_q(\frac{1}{N_f})/\Ga_q(\frac{1}{2N_f})^2$
with $q=e^{iN_f\hbar}$.
A similar expression
of instanton coefficient
has appeared in the $(1,q)$-model studied in \cite{Hatsuda:2015lpa}.

It is more interesting to determine
the coefficients of $e^{-\mu},e^{-2\mu},$ or $e^{-2\mu/N_f}$ terms,
which have non-vanishing contributions at $\hbar=2\pi$.
However, using our data of WKB expansion alone,
we were unable to find an all order expression of those terms and set
$\hbar=2\pi$.
It would be interesting  to
find those coefficients by computing the WKB expansion to more higher orders,
or by other means.

\section{WKB expansion (II)}
\label{sec:WKB2}

In the previous section, we have considered the WKB expansion
using $\rho$ satisfying $\Tr(\rho^\ell R)=0$ \cite{Assel:2015hsa}.
Alternatively, as in \cite{Mezei:2013gqa}
 we can use 
  $\rho$
  with $\Tr(\rho^\ell R)\not=0$.
In this section, we will consider the WKB expansion
using the operator in the latter case.
Interestingly, as we will see below 
we find that
the $O(2N)+A$(or $S$) model
and $O(2N+1)+A$(or $S$) model
can be thought of as a $R=1$ and $R=-1$ subspace, respectively,
of some bigger model. Namely
the $O(n)$ model corresponds to a projection to
\begin{equation}
 R=\varepsilon,
\end{equation}
depending on the parity $\varepsilon=(-1)^n$
of $n$ in \eqref{signep}.

 Using the relations
 \begin{equation}
  \bra x|\frac{1+R}{2\cosh\frac{\h{p}}{2}}|y\ket
    =\frac{1}{2\pi}\frac{4\cosh\frac{x}{2}\cosh\frac{y}{2}}{2\cosh x+2\cosh y},\qquad
    \bra x|\frac{1-R}{2\cosh\frac{\h{p}}{2}}|y\ket
    =\frac{1}{2\pi}\frac{4\sinh\frac{x}{2}\sinh\frac{y}{2}}{2\cosh x+2\cosh y},
 \end{equation}
 one can show that the density matrices in \eqref{rhoxy}
 correspond to the following quantum mechanical operators
$\rho=e^{-H}$
  \begin{equation}
\rho=\left\{
\begin{aligned}
 \rho_A\frac{1+\varepsilon R}{2}&\qquad\text{for}~~O(n)+A,\\
\rho_S\frac{1+\varepsilon R}{2}&\qquad\text{for}~~O(n)+S,
\end{aligned}\ri.
 \end{equation}
where
$\rho_A$ and $\rho_S$ are given by
\begin{equation}
  \rho_A=\frac{1}{2\cosh\frac{\h{p}}{2}}\frac{2\cosh \h{x}}{(2\cosh\frac{\h{x}}{2})^{2N_f+2}},\qquad
 \rho_S=\frac{1}{2\cosh\frac{\h{p}}{2}}\frac{(2\cosh \h{x})^{-1}}{(2\cosh\frac{\h{x}}{2})^{2N_f+2}},
\label{rhoAS}
\end{equation}
with $[\h{x},\h{p}]=2\pi i$.
Note that $\rho_{A}$ and $\rho_S$ have a reflection symmetry
\begin{equation}
 [\rho_A,R]=[\rho_S,R]=0,
\end{equation}
but $\Tr(\rho_{A,S}R)\not=0$.
It is interesting that the density matrix of $O(n)+A$ model and $O(n)+S$ model
can be written in a very similar form. 
We can treat $\rho_A$ and $\rho_S$ uniformly
by introducing a sign $\si$
\begin{equation}
 \si=\left\{\begin{aligned}
	     +1&\quad\text{for}~\rho_A,\\
	     -1&\quad\text{for}~\rho_S.
 \end{aligned}\ri.
\end{equation}
In other words,
this sign $\si$ distinguishes
the symmetric and anti-symmetric hypermultiplets.
Then \eqref{rhoAS} can be written
as
\begin{equation}
 \rho_\si=\frac{1}{2\cosh\frac{\h{p}}{2}}\frac{(2\cosh \h{x})^\si}{(2\cosh\frac{\h{x}}{2})^{2N_f+2}}.
\end{equation}
One can easily show that
the total grand potential for $\rho=\rho_\si\frac{1+\varepsilon R}{2}$
 can be decomposed as
 \begin{equation}
   J(\mu)=\frac{J_{\rho_\si}(\mu)+ \varepsilon J_{\rho_\si}^R(\mu)}{2},
\label{JinJR}
 \end{equation}
 where
 \begin{equation}
  J_{\rho_\si}(\mu)=\sum_{\ell=1}^\infty \frac{(-1)^{\ell-1}e^{\ell\mu}}{\ell}\Tr(\rho_\si^\ell),\qquad
  J_{\rho_\si}^R(\mu)=\sum_{\ell=1}^\infty \frac{(-1)^{\ell-1}e^{\ell\mu}}{\ell}\Tr(\rho_\si^\ell R).
 \end{equation}
Note that \eqref{JinJR} implies the following relation
\begin{equation}
 \begin{aligned}
  J^{O(2N)+A}(\mu)-J^{O(2N+1)+A}(\mu)&=J_{\rho_A}^R(\mu),\\
 J^{O(2N)+S}(\mu)-J^{O(2N+1)+S}(\mu)&=J_{\rho_S}^R(\mu).
 \end{aligned}
\end{equation}
The perturbative part of this relation is closely related to the difference of $B$ and $A$
found in \eqref{diff-AB}.
 From the brane configuration in section \ref{sec:fermigas},
 it is tempting to identify
 $J_{\rho_\si}^R(\mu)$ as the contribution
 of a half D2-brane stuck on the orientifold plane.
 In the rest of this section, we will consider the WKB expansion of this contribution.

 Notice that $\Tr(O R)$ for some operator $O$
can be obtained from the Wigner transform
$O_W$ by simply setting $X=P=0$\footnote{We would like to thank Yasuyuki Hatsuda for pointing this out to us.}
 \begin{equation}
  \Tr(OR)=\hf O_W\Big|_{X=P=0},
 \end{equation}
 which follows directly from the definition of $O_W$ in \eqref{Wignerdef}
\begin{equation}
 O_W\Big|_{X=P=0}=\int dy\, \Big\bra \!-\hf y\Big|O\Big|\hf y\Big\ket
=2\Tr(OR).
\end{equation}
Thus the WKB expansion of $\Tr(\rho_\si^\ell R)$ can be easily found from the
WKB expansion of $(\rho_\si^\ell)_W$,
which can be systematically computed by using the method in the previous section.
 The Wigner transform of $\rho_\si$ is
easily found to be
 \begin{equation}
  (\rho_\si)_W= \frac{1}{2\cosh\frac{P}{2}}\star \frac{(2\cosh X)^\si}{(2\cosh\frac{X}{2})^{2N_f+2}}.
 \end{equation}
 As in the previous section,
 $J_{\rho_\si}^R(\mu)$ has a Mellin-Barnes representation
 \begin{equation}
  J_{\rho_\si}^R(\mu)=-\int_\ga\frac{ds}{2\pi i}\Ga(-s)\Ga(s)\Tr(\rho_\si^sR).
   \label{BarnesJR}
 \end{equation}
We will call $\Tr(\rho_\si^sR)$
in \eqref{BarnesJR} the {\it twisted spectral trace}.
The WKB expansion of  $J_{\rho_\si}^R(\mu)$ can be found from
 the WKB expansion of twisted spectral trace $\Tr(\rho_\si^sR)$
 \begin{equation}
  \begin{aligned}
   \Tr(\rho_\si^sR)&=Z_0^R(s)D^R(s,\hbar),\nn
   D^R(s,\hbar)&=1+\sum_{n=1}^\infty D_n^R(s)\hbar^{2n}.
  \end{aligned}
  \label{WKB-R}
 \end{equation}
 The leading term $Z_0^R(s)$ in \eqref{WKB-R} can be easily obtained  as
 \begin{equation}
Z_0^R(s)=2^{-1-s(2N_f+3-\si)}.
\label{Z0R}
 \end{equation}
Note that the leading term $Z_0^R(s)$
of twisted spectral trace $\Tr(\rho_\si^sR)$
in \eqref{Z0R} is order $\mathcal{O}(\hbar^0)$,
while the leading term
$Z_0(s)$ of spectral trace $\Tr(\rho_D^s)$
in \eqref{Z0} is order $\mathcal{O}(\hbar^{-1})$.
We have computed $D_n^R(s)$ in \eqref{WKB-R} up to $n=9$ for both
 $\rho_A$ and $\rho_S$\footnote{In a similar manner,
 we can compute the WKB expansion of the twisted spectral trace
 of ABJM theory $\Tr(\rho_\text{ABJM}^sR)$ \cite{YH,KO}.
This quantity $\Tr(\rho_\text{ABJM}^sR)$
plays an important role in
the study of $\mathcal{N}=5$ $O(n)\times USp(n')$ Chern-Simons-matter theories
in the Fermi gas formalism \cite{Moriyama:2015asx,Honda}.}.
  For $\rho_A$, the first three terms are given by
  \begin{equation}
   \begin{aligned}
 D_1^R(s)&=\frac{1}{64} (1-N_f) s^2,\\
D_2^R(s)&=s^2 \left(\frac{5 (N_f-1)^2
   s^2}{24576}+\frac{(N_f+1) (2 N_f-5)
   s}{8192}+\frac{(N_f+1) (2
 N_f-5)}{12288}\right),\\
D_3^R(s)&=s^2 \left(-\frac{61 (N_f-1)^3
   s^4}{23592960}
-\frac{7 (N_f-1) (N_f+1) (2
   N_f-5) s^3}{1572864}\ri.\\
&\hskip12mm-\frac{(N_f+1) (34 N_f^2-107
   N_f-17) s^2}{2359296}
-\frac{(N_f-4)
   (N_f+1) (2 N_f+3) s}{196608}\\
&\hskip12mm\lf. -\frac{(N_f+1)
  (2 N_f^2-3
   N_f-29)}{737280}\right),
\end{aligned}
  \end{equation}
  while for $\rho_S$, the first three terms are given by
  \begin{equation}
\begin{aligned}
 D_1^R(s)&=-\frac{1}{64} (N_f+3) s^2,\\
D_2^R(s)&=s^2 \left(\frac{5 (N_f+3)^2
   s^2}{24576}+\frac{(2 N_f^2+13 N_f+27)
   s}{8192}+\frac{2 N_f^2+13
   N_f+27}{12288}\right),\\
D_3^R(s)&=s^2 \left(-\frac{61
   (N_f+3)^3 s^4}{23592960}
-\frac{7 (N_f+3) (2 N_f^2+13
   N_f+27) s^3}{1572864}\ri.\\
&\hskip12mm+\frac{(-34
   N_f^3-335 N_f^2-1272 N_f-1899)
   s^2}{2359296}+\frac{(-2 N_f^3-21
   N_f^2-91 N_f-168) s}{196608}\\
&\hskip12mm\lf.-\frac{(N_f+5)
   (2 N_f^2+13
   N_f+51)}{737280}\right).
\end{aligned}   
  \end{equation}

  As discussed in the previous section,
  deforming the contour in the direction $\Re(s)\leq0$
  in \eqref{BarnesJR},
  we can find the large $\mu$
  expansion of $J_{\rho_\si}^R(\mu)$.
  The perturbative part comes from the pole at $s=0$.
  Since the higher order terms $D_n^R(s)$ has a zero
 at $s=0$ of order $s^2$,
 they do not contribute
 to the residue at $s=0$.
 Thus, 
 the residue at $s=0$
 comes only from the leading term $Z_0(s)$
 \begin{equation}
 -\oint_{s=0} \frac{ds}{2\pi i}\Gamma(-s)\Ga(s)Z_0^R(s)e^{s\mu}=\frac{\mu}{2}-\frac{2N_f+3-\si}{2}\log2.
\label{oints-=0}
 \end{equation}
   One can see that \eqref{oints-=0} precisely reproduces the difference of $B$
 and $A$ in \eqref{diff-AB}
 between the $O(2N)$ and $O(2N+1)$ models.

 \subsection{Comments on the non-perturbative corrections}
  By matching the WKB expansion
  of twisted spectral trace $\Tr(\rho_\si^s R)$, we find the
  first two non-perturbative corrections,
coming from the poles at $s=-1$ and $s=-2$,
  in a closed form in $\hbar$
 \begin{equation}
\begin{aligned}
 &-\int \frac{ds}{2\pi i}\Gamma(-s)\Ga(s)\Tr(\rho_\si^s R)e^{s\mu}\\
=&\frac{\mu}{2}-\frac{2N_f+3-\si}{2}\log2
+\frac{(2\cos\frac{\hbar}{8})^{2(N_f+1)}}{(2\cos\frac{\hbar}{4})^\si}e^{-\mu}-\frac{(4\cos\frac{\hbar}{4})^{2(N_f+1)}}{(4\cos\frac{\hbar}{2})^\si}e^{-2\mu}+\mathcal{O}(e^{-3\mu}).
\end{aligned}
\label{trR-inst}
 \end{equation}
 It is tempting to identify these corrections as the
 ``half instantons''.
 For $\rho_S~(\si=-1)$,
the non-perturbative corrections in \eqref{trR-inst} vanish at $\hbar=2\pi$, which is
consistent with the absence of $\mathcal{O}(e^{-\mu})$ term
in $O(n)+S$ models.

On the other hand, for $\rho_A~(\si=+1)$
the 1-instanton term has a pole at $\hbar=2\pi$
\begin{align}
 \lim_{\hbar\to2\pi}\frac{(2\cos\frac{\hbar}{8})^{2(N_f+1)}}{2\cos\frac{\hbar}{4}}e^{-\mu}
=\left[-\frac{2^{N_f+2}}{\hbar-2\pi}+(N_f+1)2^{N_f}\right]e^{-\mu}.
\label{pole-R}
\end{align}
This should be canceled by a term of order $\mathcal{O}(e^{-2\pi\mu/\hbar})$, which is 
non-perturbative  in $\hbar$ and hence cannot be seen directly
in the WKB expansion.
As discussed in \cite{Hatsuda:2015oaa}, the
$\mathcal{O}(e^{-2\pi\mu/\hbar})$ term might arise from a pole
of the twisted spectral trace $\Tr(\rho_A^s R)$
at $s=-2\pi/\hbar$.
Our result of 1-instanton and 2-instanton in \eqref{trR-inst}
suggests that $\Tr(\rho_A^s R)$ has a structure
\begin{align}
 \Tr(\rho_A^s R)=\frac{f(s,N_s)}{\cos\frac{s\hbar}{4}},
\end{align}
which has a pole at $s=-2\pi/\hbar$.
As in \cite{Hatsuda:2015oaa},
using the Pade approximation,
we have checked numerically that $\Tr(\rho_A^s R)$
has a pole very close to
$s=-2\pi/\hbar$.
For the special value of $N_f=-1$, which is not physical though,
by matching the WKB expansion
we  find a closed form expression of
the twisted spectral trace $\Tr(\rho_A^s R)$
\begin{align}
 \Tr(\rho_A^s R)\Big|_{N_f=-1}=\frac{1}{2\cos\frac{s\hbar}{4}},
 \label{Nf-1}
\end{align} 
which indeed has a pole at $s=-2\pi/\hbar$, as expected.
Although we do not have an analytic proof that
$\Tr(\rho_A^s R)$ has a pole at $s=-2\pi/\hbar$
for general $N_f$,
we will assume that this is the case in the rest of this section.

We assume that $\Tr(\rho_A^s R)$
has a simple pole at $s=-2\pi/\hbar$
for general $N_f$
\begin{align}
 \lim_{s\to-\frac{2\pi}{\hbar}}\Tr(\rho_A^s R)=\frac{1}{\hbar}\frac{r(\frac{2\pi}{\hbar})}{s+\frac{2\pi}{\hbar}}.
\end{align}
Then the contribution of pole at $s=-2\pi/\hbar$
to $J_{\rho_A}^R$
is
given by
\begin{align}
 -\oint_{s=-\frac{2\pi}{\hbar}}\frac{ds}{2\pi i}\Ga(-s)\Ga(s)\Tr(\rho_A^s R)e^{s\mu}=\frac{r(\frac{2\pi}{\hbar})}{2\sin\frac{2\pi^2}{\hbar}}e^{-\frac{2\pi\mu}{\hbar}}.
\end{align}
This contribution has a pole at $\hbar=2\pi$, 
and  behaves in the limit $\hbar\to2\pi$ as
\begin{align}
 \lim_{\hbar\to2\pi}\frac{r(\frac{2\pi}{\hbar})}{2\sin\frac{2\pi^2}{\hbar}}e^{-\frac{2\pi\mu}{\hbar}}
=\left[\frac{r(1)}{\hbar-2\pi}+
\frac{(\mu+1)r(1)-r'(1)}{2\pi}\right]e^{-\mu}.
\label{pole-h}
\end{align}
Thus, there is a possibility that
the pole at $\hbar=2\pi$ cancels between
\eqref{pole-R} and \eqref{pole-h}.
This pole cancellation occurs if
$r(1)$ is given by
\begin{align}
 r(1)=2^{N_f+2}.
\end{align}
If we further assume
\begin{align}
 r'(1)=\pi(N_f+1)2^{N_f+1},
\end{align}
then the total contribution correctly reproduces the
1-half instanton term in $J_{\text{half}}$ \eqref{half-inst}
\begin{align}
 \frac{\varepsilon}{2} J_{\rho_A}^R=
 \varepsilon\frac{(\mu+1)}{\pi}2^{N_f}e^{-\mu}+\mathcal{O}(e^{-2\mu}).
\end{align}
For the $N_f=-1$ case in \eqref{Nf-1},
one can see that the residue at $s=-2\pi/\hbar$
indeed satisfies the
above conditions
\begin{align}
 r=2,\qquad r'=0.
\end{align}
It would be interesting to
study the analytic structure of $\Tr(\rho_A^s R)$
for general $N_f$ and see if
it indeed has a pole at $s=-2\pi/\hbar$
with the correct residue.

\section{Conclusion}
\label{sec:conclusion}

In this paper, we have studied non-perturbative
effects in $\mathcal{N}=4$ $O(n)$ Yang-Mills theories
with $N_f$ fundamental and one (anti)symmetric hypermultiplets
using the Fermi gas formalism
for their $S^3$ partition functions.
They are a natural generalization of $N_f$ matrix model
and interesting in their own right since 
we can study the effects of orientifold plane
in the strong coupling M-theoretic regime.

We determined the coefficients
$C,B$ and $A$ in the perturbative part of grand potential
as functions of $N_f$ in Table \ref{table:ABC}.
We also studied instanton corrections to
the grand potential using
our exact values of canonical partition functions
$Z(N,N_f)$.
We found that instanton corrections
in the $O(n)+A$ model and the $O(n)+S$ model
have slightly different structure.
For the $O(n)+A$ model,
in addition to the worldsheet instantons
\eqref{WS-O+A} and membrane instantons \eqref{M2-cand},
there are 
``half instanton'' corrections coming from the effect
of orientifold plane.
We have argued that
half instantons can be naturally identified
as the contributions from
the twisted spectral trace $\Tr(\rho_A^s R)$
in the Fermi gas picture.
Also, we found a bound state of worldsheet instanton
and half instanton
\eqref{bound}.
From
the pole cancellation argument,
there should also be a bound state of membrane instanton
and half instanton as well.

It is interesting that the reflection $R$
of one-dimensional Fermi gas system has a relation to
the orientifolding
in the spacetime.
We find that the half instanton has a weight $e^{-\mu}$
which is half of the weight of membrane instanton
$e^{-2\mu}$.
This type of half instanton
corrections is also observed in other 
theories \cite{Grassi:2014uua,Moriyama:2015asx},
and we believe that this is a general
phenomenon in M-theory on orientifolds.
It would be interesting to study
the general structure of half instantons and
clarify the precise relation to the Type IIA brane picture.

In the case of ABJM theory,
the effect of bound state can be removed
by introducing the effective chemical potential 
given by the quantum A-period \cite{HMO-bound,HMMO}.
It would be very interesting to
see whether a similar redefinition of
chemical potential works
in our case.
To study instanton corrections further,
it is desirable to
find a systematic method to
compute the WKB expansion.
In the case of $N_f$ matrix model, it has a natural
one-parameter generalization of the model by introducing a Chern-Simons level
$k$, and we can systematically study the WKB expansion
around $k=0$ \cite{Hatsuda:2015lpa}. 
It would be interesting to
find  a generalization of
the $O(n)+A$(or $S$) models
with Chern-Simons terms, along the lines of
\cite{Moriyama:2015jsa,Assel:2015hsa}.

\vskip7mm
\centerline{\bf Acknowledgments}
\noindent
I am grateful to Yasuyuki Hatsuda, Masazumi Honda,
and Marcos Marino for useful discussions.
I would also like to thank the theory group in
University of Geneva for hospitality.

\vskip6mm
\appendix

\section{$J_\text{np}(\mu,N_f)$ for various $N_f$}
\label{se:Jnp}
In this Appendix, we summarize the
non-perturbative corrections for various (half-)integral
$N_f$, obtained from the data of
exact values of $Z(N,N_f)$.
\subsection{$O(2N)+A$}\label{se:O-even+A}
Here we summarize the non-perturbative corrections $J_\text{np}(\mu,N_f)$
to the grand potential
of the $O(2N)+A$ model.

For integer $N_f$ we find
\begin{align}
 J_\text{np}(\mu,1)&=\frac{2(\mu+1)}{\pi}e^{-\mu}
+\left[-\frac{10\mu^2+7\mu+7/2}{\pi^2}+1\right]e^{-2\mu}
+\frac{88\mu+52/3}{3\pi}e^{-3\mu}\nn
J_\text{np}(\mu,2)&=\frac{2(\mu+1)}{\pi}e^{-\mu}
+\left[-\frac{39\mu^2+63\mu/2+63/4}{\pi^2}+22\right]e^{-2\mu}
+\frac{664\mu+484/3}{3\pi}e^{-3\mu}\nn
J_\text{np}(\mu,3)&=-\frac{5}{\rt{3}\pi}\left(\frac{2\mu}{3}+1\right)e^{-\frac{2\mu}{3}}
+\frac{8(\mu+1)}{\pi}e^{-\mu}
+\left[-\frac{50(\mu+3/2)^2}{9\pi^2}+\frac{100}{27}\right]e^{-\frac{4\mu}{3}}
+\frac{32}{\rt{3}}e^{-\frac{5\mu}{3}}
\nn
J_\text{np}(\mu,4)&=-\frac{6}{\rt{2}\pi}\left(\frac{\mu}{2}+1\right)e^{-\frac{\mu}{2}}
+\left[-\frac{9(\mu+2)^2}{2\pi^2}
+\frac{20(\mu+1)}{\pi}+\frac{9}{4}\right]e^{-\mu}\nn
J_\text{np}(\mu,6)&=
-\frac{8}{\pi}\left(\frac{\mu}{3}+1\right)e^{-\frac{\mu}{3}}
+\left[-\frac{32(\mu+3)^2}{9\pi^2}+\frac{110(\mu+3/2)}{9\rt{3}\pi}\right]e^{-\frac{2\mu}{3}}
\label{O-even-A-intNf}
\end{align}
and for half-integer $N_f$ we find
\begin{align}
 J_\text{np}(\mu,1/2)&=\frac{\rt{2}(\mu+1)}{\pi}e^{-\mu}
 +\left[-\frac{2(\mu+1)^2}{\pi^2}+\qu\right]e^{-2\mu}
 +\left[\frac{8\rt{2}(\mu+1)^3}{\pi^3}-\frac{\rt{2}(\mu+1)}{\pi}\right]e^{-3\mu}\nn
J_\text{np}(\mu,3/2)&=\frac{2\rt{2}(\mu+1)}{\pi}e^{-\mu}
-\frac{7}{2\rt{3}\pi}\left(\frac{4\mu}{3}+1\right)e^{-\frac{4\mu}{3}}
+\left[-\frac{8(\mu+1)^2}{\pi^2}+3\right]e^{-2\mu}
+\frac{8\rt{2}}{\rt{3}}e^{-\frac{7\mu}{3}}\nn
 &+\left[-\frac{49}{8\pi^2}\left(\frac{4\mu}{3}+1\right)^2+\frac{196}{27}\right]e^{-\frac{8\mu}{3}}+2^{\frac{9}{2}}\left[\frac{4(\mu+1)^3}{3\pi^3}-\frac{3}{2}\frac{\mu+1}{\pi}\right]e^{-3\mu}
 -\frac{64(\mu+1)}{\rt{3}\pi}e^{-\frac{10\mu}{3}}\nn
J_\text{np}(\mu,5/2)&=J_{\text{WS}}^{(1)}(\mu,5/2)
+\frac{4\rt{2}(\mu+1)}{\pi}e^{-\mu}+J_{\text{WS}}^{(2)}(\mu,5/2)
 +\frac{2^{\frac{5}{2}+1}}{\sin\frac{2\pi}{5}}e^{-\frac{9\mu}{5}}
+\left[-\frac{32(\mu+1)^2}{\pi^2}+20\right]e^{-2\mu}
\label{O-even-A-halfNf}
\end{align}
where $J^{(n)}_\text{WS}(\mu,N_f)$ is the worldsheet $n$-instanton contribution
given by \eqref{WS-O+A}.
\subsection{$O(2N+1)+A$}\label{se:O-odd+A}
Here we summarize the non-perturbative corrections $J_\text{np}(\mu,N_f)$
to the grand potential
of the $O(2N+1)+A$ model.

For integer $N_f$ we find
\begin{align}
 J_\text{np}(\mu,1)&=-\frac{2(\mu+1)}{\pi}e^{-\mu}+\left[-\frac{10\mu^2+7\mu+7/2}{\pi^2}+1\right]e^{-2\mu}
-\frac{88\mu+52/3}{3\pi}e^{-3\mu}\nn
J_\text{np}(\mu,2)&=-\frac{6(\mu+1)}{\pi}e^{-\mu}+\left[-\frac{39\mu^2+63/2\mu+63/4}{\pi^2}+6\right]e^{-2\mu}
-\frac{664\mu+484/3}{\pi}e^{-3\mu}\nn
J_\text{np}(\mu,3)&=-\frac{5}{\rt{3}\pi}\left(\frac{2\mu}{3}+1\right)e^{-\frac{2\mu}{3}}
-\frac{8(\mu+1)}{\pi}e^{-\mu}
+\left[-\frac{50(\mu+3/2)^2}{9\pi^2}+\frac{100}{27}\right]e^{-\frac{4\mu}{3}}
-\frac{32}{\rt{3}}e^{-\frac{5\mu}{3}}
\nn
J_\text{np}(\mu,4)&=-\frac{6}{\rt{2}\pi}\left(\frac{\mu}{2}+1\right)e^{-\frac{\mu}{2}}
+\left[-\frac{9(\mu+2)^2}{2\pi^2}-\frac{12(\mu+1)}{\pi}+\frac{9}{4}\right]e^{-\mu}\nn
J_\text{np}(\mu,6)&=-\frac{8}{\pi}\left(\frac{\mu}{3}+1\right)e^{-\frac{\mu}{3}}
+\left[-\frac{32(\mu+3)^2}{9\pi^2}+\frac{110(\mu+3/2)}{9\rt{3}\pi}\right]e^{-\frac{2\mu}{3}}
\label{O-odd-A-intNf}
\end{align}
and for half-integer $N_f$ we find
\begin{align}
 J_\text{np}(\mu,1/2)&=-\frac{\rt{2}}{\pi}(\mu+1)e^{-\mu}
 +\left[-\frac{2(\mu+1)^2}{\pi^2}+\qu\right]e^{-2\mu}
 +\left[-\frac{8\rt{2}(\mu+1)^3}{3\pi^3}+\frac{\rt{2}(\mu+1)}{\pi}\right]e^{-3\mu}\nn
J_\text{np}(\mu,3/2)&=-\frac{2\rt{2}}{\pi}(\mu+1)e^{-\mu}
-\frac{7}{2\rt{3}\pi}\left(\frac{4\mu}{3}+1\right)e^{-\frac{4\mu}{3}}
 +\left[-\frac{8(\mu+1)^2}{\pi^2}+3\right]e^{-2\mu} -\frac{8\rt{2}}{\rt{3}}e^{-\frac{7\mu}{3}}\nn
&+\left[-\frac{49}{8\pi^2}\left(\frac{4\mu}{3}+1\right)^2
+\frac{196}{27}\right]e^{-\frac{8\mu}{3}}
+2^{\frac{9}{2}}\left[-\frac{4 (\mu+1)^3}{3\pi^3}+\frac{3(\mu+1)}{2\pi}\right]e^{-3\mu}
-\frac{64(\mu+1)}{\rt{3}\pi}e^{-\frac{10\mu}{3}}\nn
 J_\text{np}(\mu,5/2)&=J_\text{WS}^{(1)}(\mu,5/2)-\frac{4\rt{2}}{\pi}(\mu+1)e^{-\mu}
 +J_\text{WS}^{(2)}(\mu,5/2)
-\frac{2^{\frac{5}{2}+1}}{\sin\frac{2\pi}{5}}e^{-\frac{9\mu}{5}}+\left[-\frac{2^5(\mu+1)^2}{\pi^2}+20\right]e^{-2\mu}
\label{O-odd-A-halfNf}
\end{align}
where $J^{(n)}_\text{WS}(\mu,N_f)$ is the worldsheet $n$-instanton contribution
given by \eqref{WS-O+A}.

\subsection{$O(2N)+S$}
\label{se:O-even+S}
Here we summarize the non-perturbative corrections $J_\text{np}(\mu,N_f)$
to the grand potential
of the $O(2N)+S$ model.

For integer $N_f$ we find
\begin{align}
 J_\text{np}(\mu,-1)&=\left[\frac{4\mu^2+2\mu+1}{2\pi^2}+2\right]e^{-2\mu}
+\left[-\frac{52\mu^2+\mu+9/4}{4\pi^2}-14\right]e^{-4\mu}
\nn
&+\left[\frac{386\mu^2-152\mu/3+77/9}{3\pi^2}+\frac{416}{3}\right]e^{-6\mu}
\nn
 J_\text{np}(\mu,0)&=\frac{4\mu^2+2\mu+1}{4\pi^2}e^{-2\mu}
+\left[-\frac{52\mu^2+\mu+9/4}{8\pi^2}+2\right]e^{-4\mu}
\nn
&+\left[\frac{386\mu^2-152\mu/3+77/9}{6\pi^2}-32\right]e^{-6\mu}
\nn
 J_\text{np}(\mu,1)&=-\frac{1}{\rt{3}\pi}\left(\frac{2\mu}{3}+1\right)e^{-\frac{2\mu}{3}}
+\left[-\frac{2(\mu+3/2)^2}{9\pi^2}+\frac{4}{27}\right]e^{-\frac{4\mu}{3}}\nn
 J_\text{np}(\mu,2)&=-\frac{2}{\rt{2}\pi}\left(\frac{\mu}{2}+1\right)e^{-\frac{\mu}{2}}
+\left[-\frac{(\mu+2)^2}{2\pi^2}+\qu\right]e^{-\mu}\nn
J_\text{np}(\mu,3)&=-\frac{3}{2\pi\sin\frac{\pi}{5}}\left(\frac{2\mu}{5}+1\right)e^{-\frac{2\mu}{5}}\nn
&+\left[-\frac{9}{2\pi^2}\left(\frac{2\mu}{5}+1\right)^2+
\frac{11}{40\pi\sin\frac{\pi}{5}}\left(\frac{4\mu}{5}+1\right)
+\frac{9}{50\sin^2\frac{2\pi}{5}}
\right]e^{-\frac{4\mu}{5}}\nn
J_\text{np}(\mu,4)&=-\frac{4}{\pi}\left(\frac{\mu}{3}+1\right)e^{-\frac{\mu}{3}}
+\left[-\frac{8(\mu+3)^2}{9\pi^2}+\frac{14(\mu+3/2)}{9\rt{3}\pi}\right]e^{-\frac{2\mu}{3}}
\end{align}
and for half-integer $N_f$ we find
\begin{align}
J_\text{np}(\mu,-3/2)&=\frac{3(2\mu+1)}{2\pi}e^{-2\mu}+\left[-\frac{10\mu^2+7\mu/2+7/8}{\pi^2}-\qu\right]e^{-4\mu}
+\frac{44\mu+13/3}{\pi}e^{-6\mu}\nn
 J_\text{np}(\mu,-1/2)&=\frac{1}{2\rt{3}\pi}\left(\frac{4\mu}{3}+1\right)e^{-\frac{4\mu}{3}}
+J_\text{WS}^{(2)}(\mu,-1/2)\nn
J_\text{np}(\mu,1/2)&=J_\text{WS}^{(1)}(\mu,1/2)
+J_\text{WS}^{(2)}(\mu,1/2)
\end{align}
where $J^{(n)}_\text{WS}(\mu,N_f)$
denotes the worldsheet $n$-instanton term in \eqref{WS-O+S}.

Note that the negative $N_f$ cases in the above list are unphysical for the $O(2N)+S$ theory.
Nonetheless, the integral defining the partition functions 
are well-defined for those values of negative $N_f$.

\subsection{$O(2N+1)+S$}
\label{se:O-odd+S}
Here we summarize the non-perturbative corrections $J_\text{np}(\mu,N_f)$
to the grand potential
of the $O(2N+1)+S$ model.

For integer $N_f$ we find
\begin{align}
 J_\text{np}(\mu,-1)&=\left[\frac{4\mu^2+2\mu+1}{2\pi^2}-2\right]e^{-2\mu}
+\left[-\frac{52\mu^2+\mu+9/4}{4\pi^2}+18\right]e^{-4\mu}\nn
&+\left[\frac{386\mu^2-152\mu/3+77/9}{3\pi^2}-\frac{608}{3}\right]e^{-6\mu}
\nn
 J_\text{np}(\mu,0)&=\frac{4\mu^2+2\mu+1}{4\pi^2}e^{-2\mu}
+\left[-\frac{52\mu^2+\mu+9/4}{8\pi^2}+2\right]e^{-4\mu}
\nn
&+\left[\frac{386\mu^2-152\mu/3+77/9}{6\pi^2}-32\right]e^{-6\mu}
\nn
 J_\text{np}(\mu,1)&=-\frac{1}{\rt{3}\pi}\left(\frac{2\mu}{3}+1\right)e^{-\frac{2\mu}{3}}
+\left[-\frac{2(\mu+3/2)^2}{9\pi^2}+\frac{4}{27}\right]e^{-\frac{4\mu}{3}}\nn
 J_\text{np}(\mu,2)&=-\frac{2}{\rt{2}\pi}\left(\frac{\mu}{2}+1\right)e^{-\frac{\mu}{2}}
+\left[-\frac{(\mu+2)^2}{2\pi^2}+\qu\right]e^{-\mu}\nn
J_\text{np}(\mu,3)&=-\frac{3}{2\pi\sin\frac{\pi}{5}}\left(\frac{2\mu}{5}+1\right)e^{-\frac{2\mu}{5}}\nn
&+\left[-\frac{9}{2\pi^2}\left(\frac{2\mu}{5}+1\right)^2+
\frac{11}{40\pi\sin\frac{\pi}{5}}\left(\frac{4\mu}{5}+1\right)
+\frac{9}{50\sin^2\frac{2\pi}{5}}
\right]e^{-\frac{4\mu}{5}}\nn
J_\text{np}(\mu,4)&=-\frac{4}{\pi}\left(\frac{\mu}{3}+1\right)e^{-\frac{\mu}{3}}
+\left[-\frac{8(\mu+3)^2}{9\pi^2}+\frac{14(\mu+3/2)}{9\rt{3}\pi}\right]e^{-\frac{2\mu}{3}}
\end{align}
and for half-integer $N_f$ we find
\begin{align}
J_\text{np}(\mu,-3/2)&=-\frac{2\mu+1}{2\pi}e^{-2\mu} +\left[-\frac{10\mu^2+7\mu/2+7/8}{\pi^2}+\frac{7}{4}\right]e^{-4\mu}
-\frac{44\mu+13/3}{3\pi}e^{-6\mu}\nn
J_\text{np}(\mu,-1/2)&=\frac{1}{2\rt{3}\pi}\left(\frac{4\mu}{3}+1\right)e^{-\frac{4\mu}{3}}
+J_\text{WS}^{(2)}(\mu,-1/2)\nn
J_\text{np}(\mu,1/2)&=-\frac{1}{4\pi\sin\frac{2\pi}{5}}\left(\frac{4\mu}{5}+1\right)e^{-\frac{4\mu}{5}}
+J_\text{WS}^{(2)}(\mu,1/2)
\end{align}
where $J^{(n)}_\text{WS}(\mu,N_f)$
denotes the worldsheet $n$-instanton term in \eqref{WS-O+S}.

Note that the negative $N_f$ cases in the above list are unphysical for the $O(2N+1)+S$ theory,
but by using the relation \eqref{O-Sp-relation}
they can be regarded as the $USp(2N)+S$ theory
with positive $N_f$.

\section{Computation of Wigner transform in \eqref{rhoDW}}
\label{se:Wigner}
In this Appendix, we will derive \eqref{rhoDW} for the Wigner transform 
of $\rho_D$ in \eqref{rho-D}.
Using the property of Wigner transformation
\begin{equation}
 f(\h{x})_W=f(X),\qquad
g(\h{p})_W=g(P),
\end{equation} 
and the star-product formula in \eqref{Wprod},
we find
\begin{equation}
 (\rho_D)_W=\frac{2\sinh\frac{X}{2}}{(2\cosh\frac{X}{2})^{2N_f+3}}\star O_W
\label{comp1}
\end{equation}
where $O$ is the operator inside the parenthesis in \eqref{rho-D}
\begin{equation}
 O=\sinh\frac{\h{x}}{2} \frac{1}{\cosh\frac{\h{p}}{2}}\cosh\frac{\h{x}}{2}+
\cosh\frac{\h{x}}{2} \frac{1}{\cosh\frac{\h{p}}{2}}\sinh\frac{\h{x}}{2}.
\end{equation}
Note that $[\h{x},\h{p}]=i\hbar$ and $\hbar=2\pi$.
To compute $O_W$, we need the matrix element of $(\cosh\frac{\h{p}}{2})^{-1}$
\begin{equation}
 \bra x|\frac{1}{\cosh\frac{\h{p}}{2}}|y\ket=\frac{1}{2\pi} \frac{1}{\cosh\frac{x-y}{2}}.
\end{equation}
Then, from the definition of $O_W$ in \eqref{Wignerdef} we find
\begin{equation}
 \begin{aligned}
  O_W&=\int \frac{dy}{2\pi}e^{\frac{iPy}{2\pi}}
\frac{\sinh (\frac{X}{2}-\frac{y}{4})\cosh (\frac{X}{2}+\frac{y}{4})+
\cosh (\frac{X}{2}-\frac{y}{4})\sinh (\frac{X}{2}+\frac{y}{4})}{\cosh \frac{y}{2}}\\
&=\int \frac{dy}{2\pi}e^{\frac{iPy}{2\pi}}\frac{\sinh X}{\cosh \frac{y}{2}}=\frac{\sinh X}{\cosh \frac{P}{2}}.
 \end{aligned}
\label{comp2}
\end{equation}
From \eqref{comp1} and \eqref{comp2}, the Wigner transform of $\rho_D$
is given by \eqref{rhoDW}.

\end{document}